\documentclass[12pt]{article}
\usepackage{amsmath}
\usepackage{anysize}

\marginsize{0.65in}{0.65in}{1in}{1in}

\usepackage{multirow}
\usepackage[dvips]{graphicx}
\usepackage{amsmath}
\usepackage[psamsfonts]{amssymb}
\usepackage{amsxtra}
\usepackage{threeparttable}
\usepackage{relsize}

\usepackage{stmaryrd}
\usepackage{amsfonts,amsmath,amssymb}

\usepackage{color}
\usepackage[table]{xcolor}

\usepackage{rotating}

\usepackage{cite}

\usepackage{algorithm}
\usepackage{algorithmic}
\usepackage{booktabs}
\usepackage{url}

\usepackage{colortbl}

\definecolor{shadingcolor}{rgb} {.87,    0.92,    .98}

\usepackage[tight,footnotesize]{subfigure}

\begin{document}

\input{def1jml.set}

\title{\bf Log-Determinant Divergences Revisited: Alpha--Beta and Gamma Log-Det Divergences}

\author{Andrzej CICHOCKI,  Sergio CRUCES and Shun-ichi AMARI\\
       Laboratory for Advanced Brain Signal Processing, Japan \\
       and
       Systems Research Institute, Polish Academy of Science, Poland \\
Dpto de Teor\'{i}a de la Se\~{n}al y Comunicaciones, University of Seville, Spain\\
Amari Laboratory for Mathematical Neuroscience, Japan}


%


\maketitle

\abstract{In this paper, we review and extend a family of  log-det divergences  for  symmetric positive definite (SPD) matrices and discuss their fundamental properties.  We  show how to generate from parameterized Alpha-Beta (AB) and Gamma log-det divergences many well known divergences, for example, the Stein's loss, S-divergence, called also Jensen-Bregman LogDet (JBLD) divergence, the Logdet Zero (Bhattacharryya) divergence, Affine Invariant Riemannian Metric (AIRM) as well as some new divergences.  Moreover, we establish links and correspondences among many log-det divergences  and display them on alpha-beta plain for various set of parameters.  Furthermore,  this paper bridges these  divergences and shows also their links to divergences of multivariate  and multilinear normal  distributions. Closed form formulas are derived for gamma divergences of two multivariate Gaussian densities including as special cases the Kullback-Leibler,  Bhattacharryya, R\'enyi and Cauchy-Schwartz  divergences. Symmetrized versions of the log-det divergences are also discussed and reviewed. A class of divergences is extended to multiway divergences for separable  covariance (or precision) matrices.\\



\section{Introduction}

Divergences or (dis)similarity measures between symmetric positive definite (SPD) matrices  are quite important in many applications including Diffusion Tensor Imaging (DTI) segmentation, classification,  clustering, recognition, model selection,  statistical inference, data processing problems to mention a just few \cite{Amari2014}, \cite{Basseville2013}.
Furthermore there is a close connection between the divergences and the notions of entropy, information geometry and mean values \cite{Basseville2013}, \cite{Amari09}, \cite{Sra2014a}, \cite{MatrixIG-Nielsen}. The matrix divergences are closely related to the invariant geometrical properties of the manifold of probability distributions
 \cite{Amari09}, \cite{Amari2009b}, \cite{JunZhang},\cite{Amari-PAN}.
%
%
A wide class of parameterized divergences have been investigated
 and their properties have been
investigated  and some  works have been made to unify or generalize them \cite{Cich-Amari-entropy}, \cite{Cich-Cruces-Am}, \cite{Cruces-Cich-entropy}, \cite{NMTF-09}.

 The set of symmetric positive definite (SPD) matrices, especially covariance matrices plays key roles in many areas of  statistics, signal/image processing, DTI, pattern recognition and biological and social sciences \cite{Cherian-Sra13},\cite{CherianS14}, \cite{Olszewski14}. For example, the medical data produced by diffusion tensor magnetic resonance imaging (DTI-MRI) represent the covariance in a Brownian motion model of water diffusion and under some physical interpretation diffusion tensors are required to be  represented as symmetric, positive-definite matrices  which are used to track the diffusion of water molecules in the human brain, with applications such as diagnosis of  some mental disorders \cite{Cherian-Sra13}.
One of the most prevalent data analysis and signal-processing tools is the analysis covariance matrices, which has many applications in clustering and classification problems. In array processing, covariance matrices capture both the variance and correlation in multidimensional data. Often this is linked to estimate (dis)similarity measures -- divergences.
In fact, in recent years we observe an increased interest in the investigation of divergences for SPD (covariance) matrices
\cite{Amari2014},\cite{Cherian-Sra13}, \cite{Sra-NIPS12} \cite{Sra2014a}, \cite{nielsen2013jensen}, \cite{Chebbi2012} \cite{QUIC14}, \cite{Nielsen12-SM}.

The main objective of this paper is to  review and extend log-determinant (briefly log-det) divergences and to establish their links  between them and the standard divergences, especially alpha, beta and gamma divergences.
Several forms of the log-det divergence
have been given in the literature, including the  Riemannian metric, Stein's loss, S-divergence, called also Jensen-Bregman LogDet (JBLD) divergence and the symmetrized Kullback-Leibler Density Metric (KLDM) or Jeffreys KL divergence.
The properties of such divergences have been already studied and they found numerous applications, however some common theoretical properties and links between them were not investigated. In this paper, we propose  parameterized a wide class of the  log-det divergences  that may provide  more robust solutions and/or  improved  accuracy for noisy data. Moreover, we provide fundamental properties and links among wide class of divergences.
 The advantages of some selected log-det divergences include  efficiency, simplicity and
resilience to noise or outliers in addition to it being relatively easy to
calculate \cite{Cherian-Sra13}.
Moreover, the log-det divergences  between  two SPD matrices  have been shown to be robust to biases in composition that can cause problems for other similarity measures.

%
The divergences discussed in this paper are flexible because they allow us to generate   well known and often used particular divergences (for specific values of tuning parameters). Moreover, by adjusting adaptive tuning parameters, we can optimize cost functions for learning algorithms  and estimate desired parameters of a model in presence of noise and outliers. In other words, the divergences discussed in this paper can be robust with respect to outliers  and noise for some values of tuning parameters: alpha, beta and gamma.

\section{Some Preliminaries}

We will use the following notations. The symmetric positive definite matrices will be denoted as
$\bP \in \Real^{n \times n}$ and $\bQ \in \Real^{n \times n}$, which have positive eigenvalues
$\lambda_i$ (usually sorted in descending order).
 $\log(\bP),\; \det(\bP)=|\bP|, \; \tr(\bP)$ denote the logarithm, determinant and trace of the matrix $\bP$, respectively.
 We will use extensively the following basic properties of matrix logarithm, determinants, and traces:
 \be
  \log(\bP^{\alpha}) = \log((\bV \mbi \Lambda \bV^T )^{\alpha})= \bV \log(\mbi \Lambda^{\alpha}) \; \bV^T,
 \ee
 where $\log (\mbi \Lambda)$ is a diagonal matrix with logarithms of the eigenvalues of $\bP$ and $\bV \in \Real^{n \times n}$ is orthogonal matrix of the corresponding eigenvectors,
 \be
  \log (\det \; \bP) &=& \tr \log (\bP),\\
(\det  \bP)^{\alpha} &=& \det(\bP^{\alpha}), \\
 \det(\bP^{\alpha}) &=& \det(\bV \mbi \Lambda \bV^T )^{\alpha}= \det (\bV)  \det (\mbi \Lambda^{\alpha}) \det (\bV^T)
 =\prod_{i=1}^n \lambda_i^{\alpha}, \\
 \tr(\bP^{\alpha}) &=& \tr(\bV \mbi \Lambda \bV^T )^{\alpha}= \tr (\bV \bV^T  \mbi \Lambda^{\alpha} )
 = \sum_{i=1}^n \lambda^{\alpha}_i , \\
 \bP^{\alpha +\beta} &=& \bP^{\alpha} \; \bP^{\beta},\\
  (\bP^{\; \alpha})^{\;\beta} &=& \bP^{\;\alpha \;\beta}\\
  \bP^{\;0} &=& \bI,\\
 (\det  \bP)^{\alpha+\beta} &=& \det(\bP^{\alpha}) \det(\bP^{\beta}), \\
 \det  ((\bP \bQ^{-1})^{\alpha}) &=& [\det(\bP) \det(\bQ^{-1})]^{\alpha} =\det(\bP^{\alpha}) \det(\bQ^{-\alpha}), \\
 \frac{\partial}{\partial \alpha} \left( \bP^{\alpha}\right) &=&  \bP^{\alpha} \log(\bP),\\
 \frac{\partial}{\partial \alpha} \log \left[\det (\bP(\alpha))\right] &=& \tr \left( \bP^{-1}  \frac{\partial \bP}{\partial \alpha} \right),\\
 \log(\det(\bP \otimes \bQ))&=& n \log(\det \bP) + n \log( \det \bQ),\\
 \tr (\bP) -\log \det(\bP) &\geq& n.
 \ee

The dissimilarity  between two SPD matrices is called a metric if the following
 conditions hold:
 \begin{enumerate}
 \item $D(\bP\, || \, \bQ) \geq 0$, where equality holds if and only if $\bP=\bQ$ (nonnegativity and positive definiteness),

 \item
 $D(\bP\, || \, \bQ) = D( \bQ \,|| \,\bP)$  (symmetry),

 \item  $D(\bP\, || \, \bZ) \leq D(\bP\,|| \,\bQ) +
 D(\bQ \,|| \, \bZ)$ (subaddivity/triangle inequality).
 \end{enumerate}
Dissimilarities which only satisfy condition (1) are not a metric and are referred to as
(asymmetric) divergences.

\section{Basic Alpha-Beta Log-Determinant Divergence}
\label{sec2.5}


For symmetric positive definite matrices $\bP \in \Real^{n \times n}$ and $\bQ  \in \Real^{n \times n}$ (both of the same size $n \times n$), let define the following function, (which will be  considered as a  new dissimilarity measure referred  briefly to as the AB log-det divergence):
\begin{eqnarray}\label{defAB1}
    D^{(\alpha,\beta)}_{AB}({\bP}\|{\bQ})
    &=&
    \frac{1}{\alpha \beta} \log \det
     \frac{\alpha (\bP \bQ^{-1})^{\beta} + \beta (\bP \bQ^{-1})^{-\alpha}}
        {\alpha+\beta} \\ \nonumber
    \\&&\ \text{for}\ \alpha \neq 0, \;\; \beta\neq 0, \;\;\; \alpha+\beta\neq 0. \nonumber
\end{eqnarray}
This is not a symmetric divergence with respect to $\bP$ and $\bQ$ except for the case $\alpha=\beta$.\\
Using  basic properties of determinants,  we can write it in an equivalent form
\begin{eqnarray}    \label{defAB2}
    D^{(\alpha,\beta)}_{AB}({\bP}\|{\bQ})
    &=&
    \frac{1}{\alpha \beta} \log
    \displaystyle
     \frac{\det \left(\displaystyle \frac{\alpha \;(\bP \bQ^{-1})^{\alpha+\beta} + \beta \; \bI}{\alpha +\beta}\right)}
        {\det (\bP \bQ^{-1})^{\alpha}} \\ \nonumber
    \\&&\  \text{for}\ \;\; \alpha,\beta,\alpha+\beta\neq 0 \nonumber 
\end{eqnarray}
We note that using the identity $\log \det(\bP) = \tr\log(\bP)$, we can express (\ref{defAB1}) as
\begin{eqnarray}    \label{defAB3}
    D^{(\alpha,\beta)}_{AB}({\bP}\|{\bQ})
    &=&
    \frac{1}{\alpha \beta} \tr \left[ \log \left(
     \frac{\alpha (\bP \bQ^{-1})^{\beta} + \beta (\bP \bQ^{-1})^{-\alpha}}
        {\alpha+\beta}\right) \right ] \\ \nonumber
    \\&&\ \text{for}\ \;\; \alpha \neq 0, \;\; \beta\neq 0, \;\;\; \alpha+\beta\neq 0. \nonumber
\end{eqnarray}
It is interesting to observe that such a divergence has some correspondences and relationships to alpha, beta and AB-divergences discussed in our previous papers, and especially gamma divergences \cite{Cich-Cruces-Am}, \cite{Cich-Amari-entropy}, \cite{NMTF-09}, see also \cite{Gamma-Eguchi}.

Furthermore, the above defined divergence is different but related to the AB divergence for SPD matrices  defined as
\begin{eqnarray}    \label{defAB4}
    \bar D^{(\alpha,\beta)}_{AB}({\bP}\|{\bQ})
    &=&
   \frac{1}{\alpha \beta} \tr \left( \frac{\alpha}{\alpha+\beta}
   \bP^{\alpha+\beta} + \frac{\beta}{\alpha+\beta}  \; \bQ ^{\alpha+\beta} -  \bP^{\alpha} \bQ^{\beta} \right)\\
    &&\  \text{for}\ \;\; \alpha \neq 0, \;\; \beta\neq 0, \;\;\; \alpha+\beta\neq 0, \nonumber
\end{eqnarray}
which will be investigated in detail in a separated paper
(see also \cite{Amari2014},\cite{Cich-Cruces-Am} ).

It should be noted that  $D^{(\alpha,\beta)}_{AB}({\bP}\|{\bQ})$, defined in (\ref{defAB2}), can be  evaluated without need to compute inverse of SPD matrices.
 It can evaluated easily by computing
(positive) eigenvalues of the matrix $\bP \bQ^{-1}$  or it is inverse.
Since both matrices  $\bP$ and $\bQ$ (and their inverses) are SPD matrices, their eigenvalues are positive.
It can be shown that although in general matrix $\bP \bQ^{-1}$  is non symmetric, its eigenvalues are the same as those of the SPD matrix $\bQ^{-1/2}\bP \bQ^{-1/2}$, so its eigenvalues are always positive.

Taking into account the  eigenvalue decomposition:
\be
(\bP \bQ^{-1})^{\beta} = \bV \mbi \Lambda^{\beta} \; \bV^{-1},
\ee
(where $\bV$ is a nonsingular matrix, while $ \mbi \Lambda^{\beta} =\diag\{\lambda_1^{\beta},\lambda_2^{\beta},\ldots,\lambda_n^{\beta}\}$ is the diagonal matrix  with the positive eigenvalues $\lambda_i >0, \; i=1,2,\ldots,n, $ of $\bP \bQ^{-1}$), we can write
\be
 D^{(\alpha,\beta)}_{AB}({\bP}\|{\bQ})
   &=&
    \frac{1}{\alpha \beta} \log \det
     \frac{\alpha  \; \bV \mbi \Lambda^{\beta} \; \bV^{-1} + \beta \; \bV \mbi \Lambda^{-\alpha} \; \bV^{-1}}
        {\alpha+\beta} \nonumber \\
         &=&  \frac{1}{\alpha \beta} \log \left[ \det \bV \;
     \det \frac{\alpha \mbi \Lambda^{\beta}  + \beta \mbi \Lambda^{-\alpha}}
        {\alpha+\beta} \; \det \bV^{-1}\right] \nonumber \\
        &=&  \frac{1}{\alpha \beta} \log \det
     \frac{\alpha \; \mbi \Lambda^{\beta}  + \beta \; \mbi \Lambda^{-\alpha}}
        {\alpha+\beta}
        \label{DABlambda1}
\ee
Hence, after simple algebraic manipulations, we obtain
\be
 D^{(\alpha,\beta)}_{AB}({\bP}\|{\bQ})
   &=&
    \frac{1}{\alpha \beta} \log \prod_{i=1}^n
     \frac{\alpha  \lambda_i^{\beta}  + \beta  \lambda_i^{-\alpha}}
        {\alpha+\beta} \nonumber \\
         &=&  \frac{1}{\alpha \beta} \sum_{i=1}^n \log \left(\frac
     {\alpha \lambda_i^{\beta}  + \beta \lambda_i^{-\alpha}}
        {\alpha+\beta}\right), \;\; \alpha,\;\beta,\; \alpha+\beta \neq 0.
        \label{DABlambda2}
\ee

It is easy to check that $D^{(\alpha,\beta)}_{AB}({\bP}\|{\bQ}) =0$ if $\bP=\bQ$. We will show later that this function is nonnegative for any SPD matrices if alpha and beta parameters  take  both positive or negative values.

For the singular values $\alpha=0$ and/or $\beta=0$ (also $\alpha=-\beta$) the AB  log-det divergence
  (\ref{defAB1}) have to be defined as limiting cases  respectively for $\alpha \rightarrow
  0$ and/or  $\beta \rightarrow 0$.
In other words, to avoid  indeterminacy or singularity  for specific values of parameters,  the AB log-det divergence can be reformulated (extended) by continuity by applying L'H\^opital's formula to cover also the singular  values of $\alpha, \beta$.  Using the L'H\^opital's rule we found that the AB log-det divergence can be expressed or defined in explicit form as:
\be
    \label{ABdef-full}
    D^{(\alpha,\beta)}_{AB}({\bP}\|{\bQ}) &=&
    \left\{
    \begin{tabular}{ll}
    $\displaystyle  \frac{1}{\alpha \beta} \log \det
     \frac{\alpha (\bP \bQ^{-1})^{\beta} + \beta (\bQ \bP^{-1})^{\alpha}}
        {\alpha+\beta}$&  $\mbox{for} \;\; \alpha,\beta\neq0, \;\;  \alpha+\beta \neq 0$ \\
    \\
    $ \displaystyle \frac{1}{\alpha^2} \left[
     \tr \left((\bQ \bP^{-1})^{\alpha} - \bI\right) - \alpha \log \det (\bQ \bP^{-1}) \right]
         $   &$\ \text{for}\ \;\; \alpha \neq 0, \; \beta= 0 $ \\
    \\
    $\displaystyle  \frac{1}{\beta^2} \left[
     \tr \left((\bP \bQ^{-1})^{\beta} - \bI \right) - \beta \log \det (\bP \bQ^{-1}) \right] $   &$\
      \text{for}\ \;\; \alpha= 0, \; \beta\neq 0 $ \\
       \\
      $\displaystyle \frac{1}{\alpha^2}
      \log \displaystyle \frac{\det(\bP \bQ^{-1})^\alpha }{\det(\bI +\log(\bP\bQ^{-1})^\alpha)}$
      &$\
      \text{for}\ \;\; \alpha=-\beta\neq 0 $ \\\\
    $\displaystyle \frac{1}{2} \tr \log^2 (\bP \bQ^{-1}) = \frac{1}{2} ||\log(\bQ^{-1/2} \bP \bQ^{-1/2})||^2_F$ & $\ \text{for}\ \;\; \alpha, \; \beta= 0 $.
    \end{tabular}
    \right.
\ee
or equivalently after simple mathematical operations it can be expressed by eigenvalues of the matrix $\bP \bQ^{-1}$ (or its transpose),  i.e., the generalized eigenvalues computed from $\lambda_i \bQ \bv_i =\bP \bv_i$, where $\bv_i$ ($i=1,2,\ldots,n$) are corresponding generalized eigenvectors:
%
\be
    \label{ABdef-full-lambda2}
    D^{(\alpha,\beta)}_{AB}({\bP}\|{\bQ}) &=&
    \left\{
    \begin{tabular}{ll}
    $\displaystyle  \frac{1}{\alpha \beta}  \sum_{i=1}^n \log \left(\frac
     {\alpha \lambda_i^{\beta}  + \beta \lambda_i^{-\alpha}}
        {\alpha+\beta}\right) $   &$\mbox{for} \;\; \alpha,\beta\neq 0, \;\; \;\; \alpha+\beta\neq0$ \\
    \\
    $ \displaystyle \frac{1}{\alpha^2} \left[ \sum_{i=1}^n \left( \lambda_i^{-\alpha} - \log (\lambda_i^{-\alpha}) \right) -n \right]
         $   &$\ \text{for}\ \;\; \alpha \neq 0, \; \beta= 0 $ \\
    \\
    $\displaystyle  \frac{1}{\beta^2} \left[ \sum_{i=1}^n \left(\lambda_i^{\beta} - \log (\lambda_i^{\beta})\right) -n \right] $   &$\
      \text{for}\ \;\; \alpha= 0, \; \beta\neq 0 $ \\
       \\
	$ \displaystyle \frac{1}{\alpha^2}
	\left[ \sum_{i=1}^n
	\log\left( \frac{\lambda_i^{\alpha}}{1+\log \lambda_i^\alpha} \right)
	\right]
	$
   &$\
      \text{for}\ \;\; \alpha=-\beta\neq 0 $ \\\\
    $\displaystyle  \frac{1}{2}  \sum_{i=1}^n \log^2 (\lambda_i) $ & $\ \text{for}\ \;\; \alpha, \; \beta= 0 $.
    \end{tabular}
    \right.
\ee

We can prove the following Theorem (see Appendix).\\

{\bf Theorem 1} The function $D^{(\alpha,\beta)}_{AB}({\bP}\|{\bQ}) \geq 0 $ expressed by Eq. (\ref{defAB1}) is nonnegative
for any  SPD matrices with arbitrary positive eigenvalues for the following set of parameters $\alpha \geq 0$ and $\beta \geq 0$ or  $\alpha < 0$ and simultaneously $\beta < 0$ and equal zero if and only if $\bP=\bQ$.

In other words if the values of $\alpha$ and $\beta$ parameters have the same sign, the AB log-det  divergence is positive independent of distribution of eigenvalues of $\bP \bQ^{-1}$ and achieves zero if and only if all eigenvalues are equal to one.

However, if the eigenvalues  are sufficiently  close to one the AB log-det divergence is also positive for different signs of $\alpha$ and $\beta$ parameters. The conditions for positive definiteness can be formulated by the following Theorem 2:

{\bf Theorem 2} The function $D^{(\alpha,\beta)}_{AB}({\bP}\|{\bQ}) $ expressed by Eq. (\ref{ABdef-full}) is non-negative for  the set of parameters $ \alpha >0 $ and $\beta < 0$, or  $\alpha  < 0$ and $\beta>0$, if all the eigenvalues of the matrix $\bP\bQ^{-1}$ satisfy the following conditions:
\be
\lambda_i &>& \left|\frac{\beta}{\alpha}\right|^\frac{1}{\alpha+\beta} \qquad \forall i,\  \text{for}\ \alpha>0\ \text{and}\ \beta<0,
\ee
and
\be
\lambda_i &<& \left|\frac{\beta}{\alpha}\right|^\frac{1}{\alpha+\beta} \qquad \forall i,\  \text{for}\ \alpha<0\ \text{and}\ \beta>0.
\ee
When any of the eigenvalues does not satisfy these bounds, the value of the divergence should be (by definition) set to infinite.

Moreover, in the limit, when $\alpha\rightarrow -\beta$ the bounds simplifies to
\be
\lambda_i&>&e^{-1/\alpha}\quad \forall i, \ \text{}  \alpha=-\beta>0, \\
\lambda_i&<&e^{-1/\alpha}\quad \forall i, \ \text{}  \alpha=-\beta<0.
\ee
Whereas, in the limit, for $\alpha\rightarrow 0$ or for $\beta\rightarrow 0$ the bounds disappear.

The complete picture of bounds for different values of $\alpha$ and $\beta$ is shown in Fig. \ref{Fig:contour}.

Additionally, $D^{(\alpha,\beta)}_{AB}({\bP}\|{\bQ})=0$ only for $\lambda_i=1$ for $i=1,\ldots,n$, i.e., when $\bP=\bQ$.\\
The Proofs are given in the Appendices \ref{Ap-1}-\ref{Ap-3}.

\begin{figure}[!t]
\centering
\noindent
\subfigure[Lower-bounds on $\lambda_i$.]
{
\includegraphics[draft=false,width=0.47\columnwidth,angle=0,trim=0cm 0cm 0cm 0cm, clip=true]{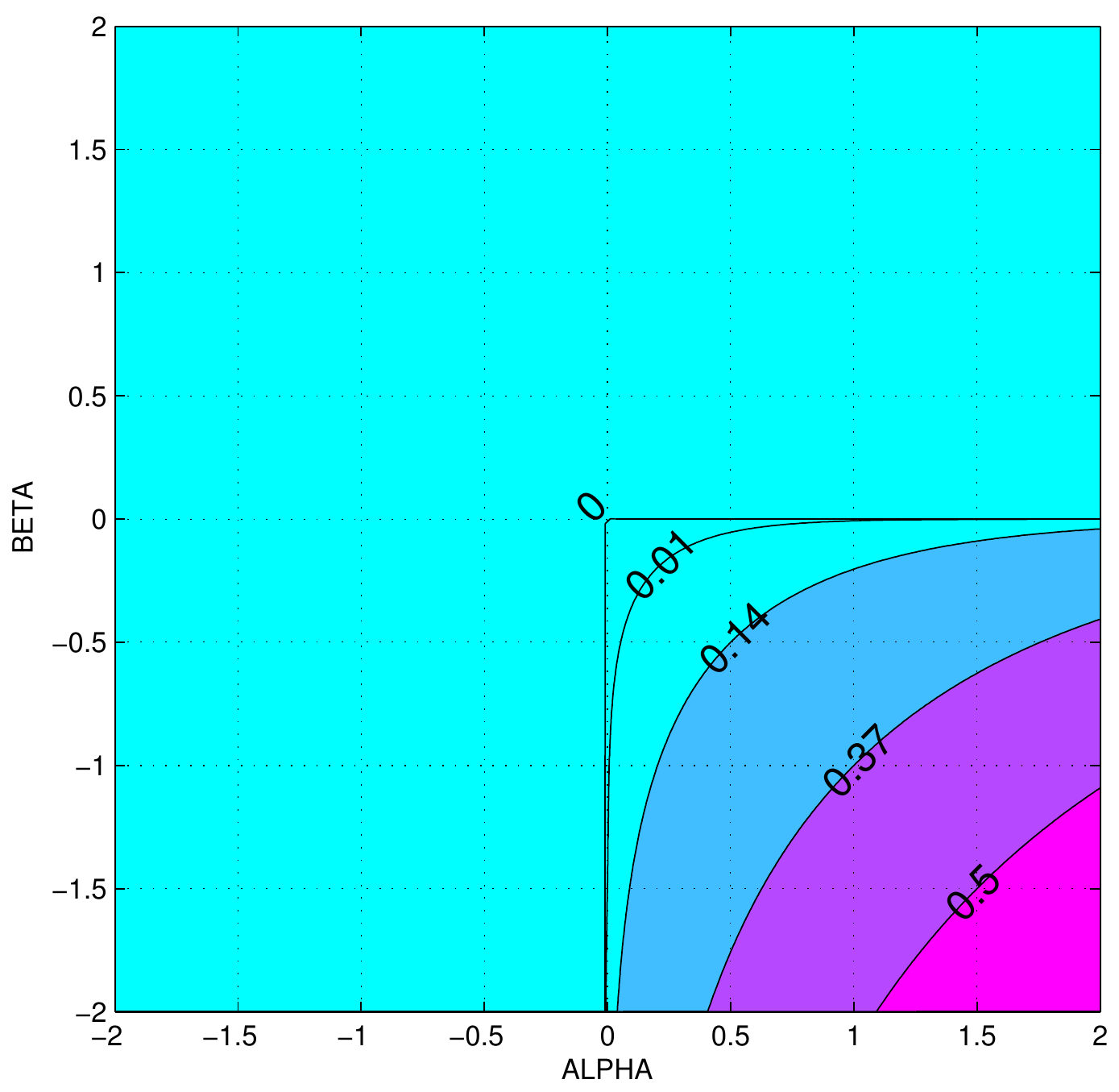}%
\label{Fig0-b}
}
\subfigure[Upper-bounds on $\lambda_i$.]
{
\includegraphics[draft=false,width=0.47\columnwidth,angle=0,trim=0cm 0cm 0cm 0cm, clip=true]{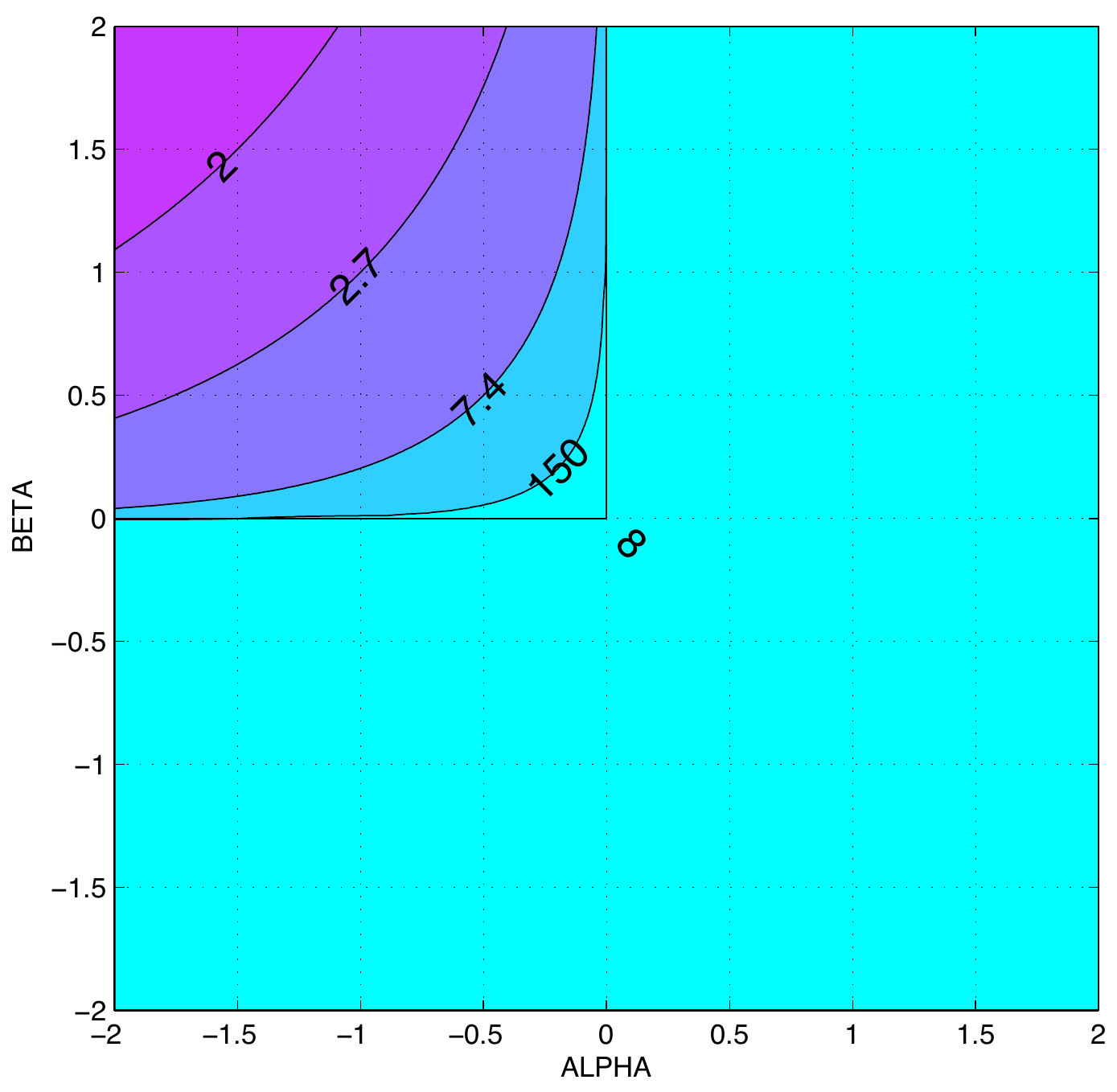}%
\label{Fig0-a}
}
\caption{Shaded-contour plots of the bounds on $\lambda_i$ that prevent $D^{(\alpha,\beta)}_{AB}({\bP}\|{\bQ})$ form diverging to $\infty$. The positive lower-bounds are in the lower-right quadrant of subfigure (a). The finite upper-bounds are in the upper-left quadrant of subfigure (b).}
\label{Fig:contour}
\end{figure}

\begin{figure}[ht!]
(a)
\includegraphics[width=8.1cm]{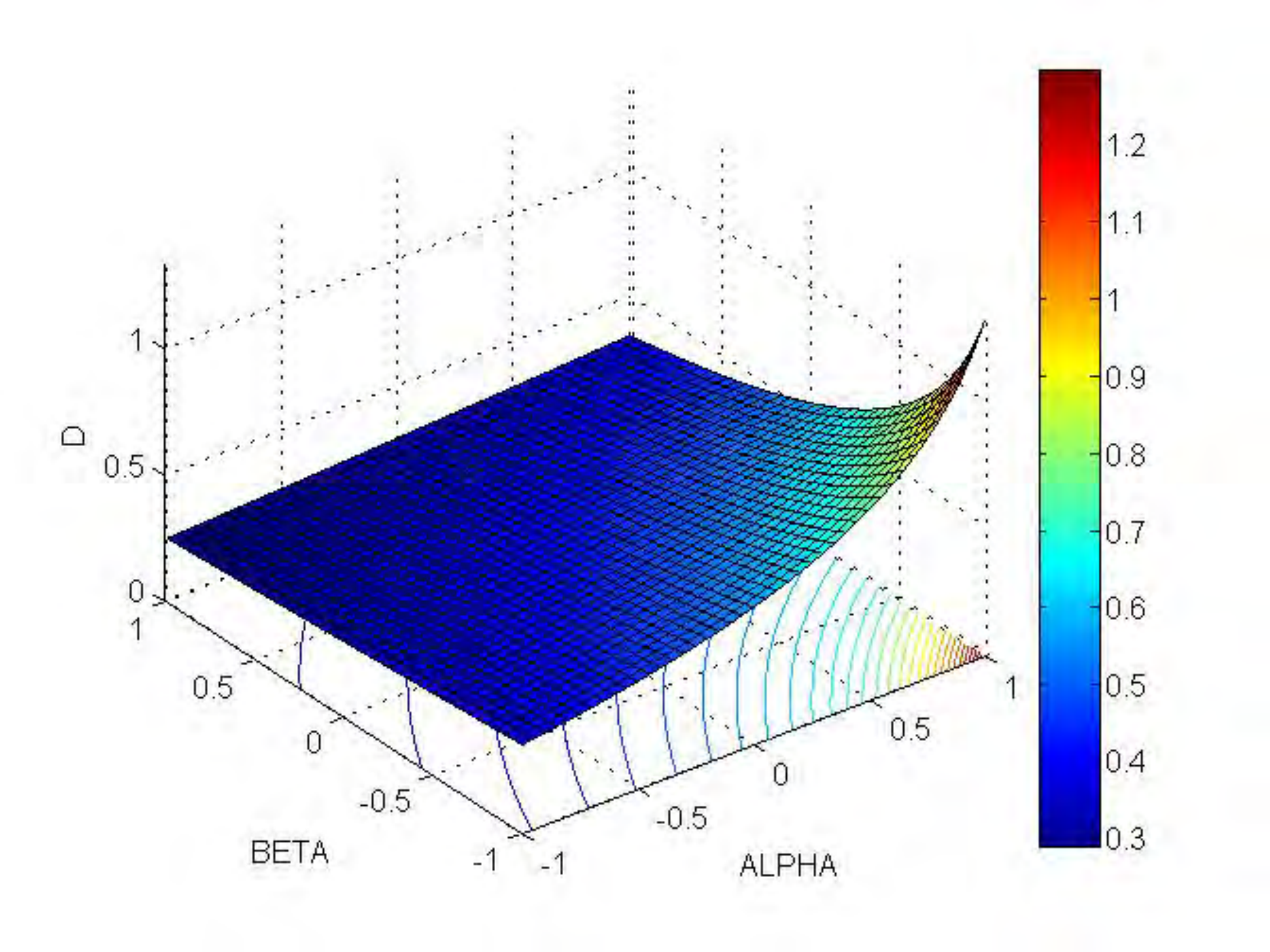}
\hspace{0.01cm}
(b)
\includegraphics[width=8.1cm]{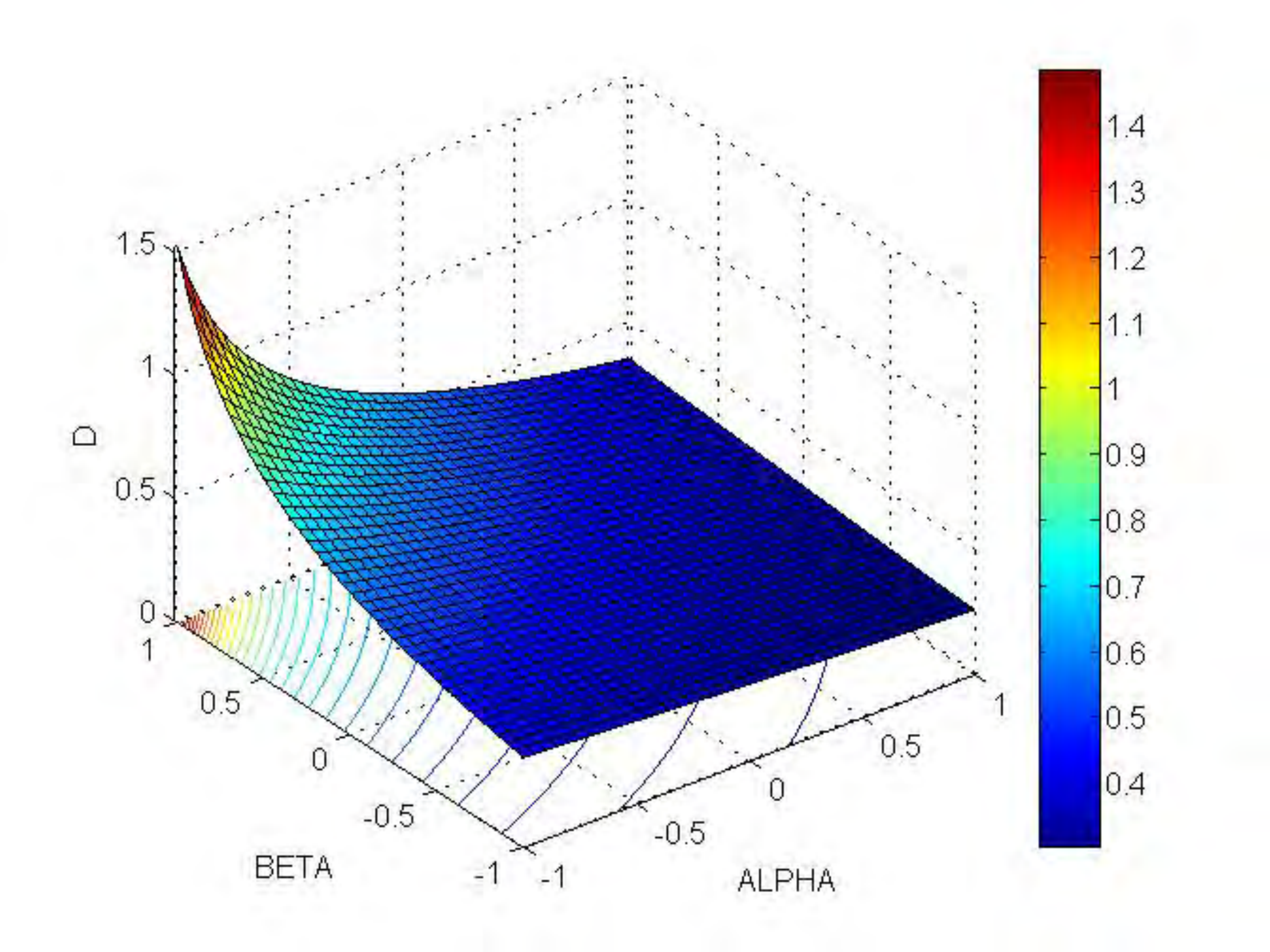}\\
\vspace{0.3cm}
(c)
\includegraphics[width=7.0cm]{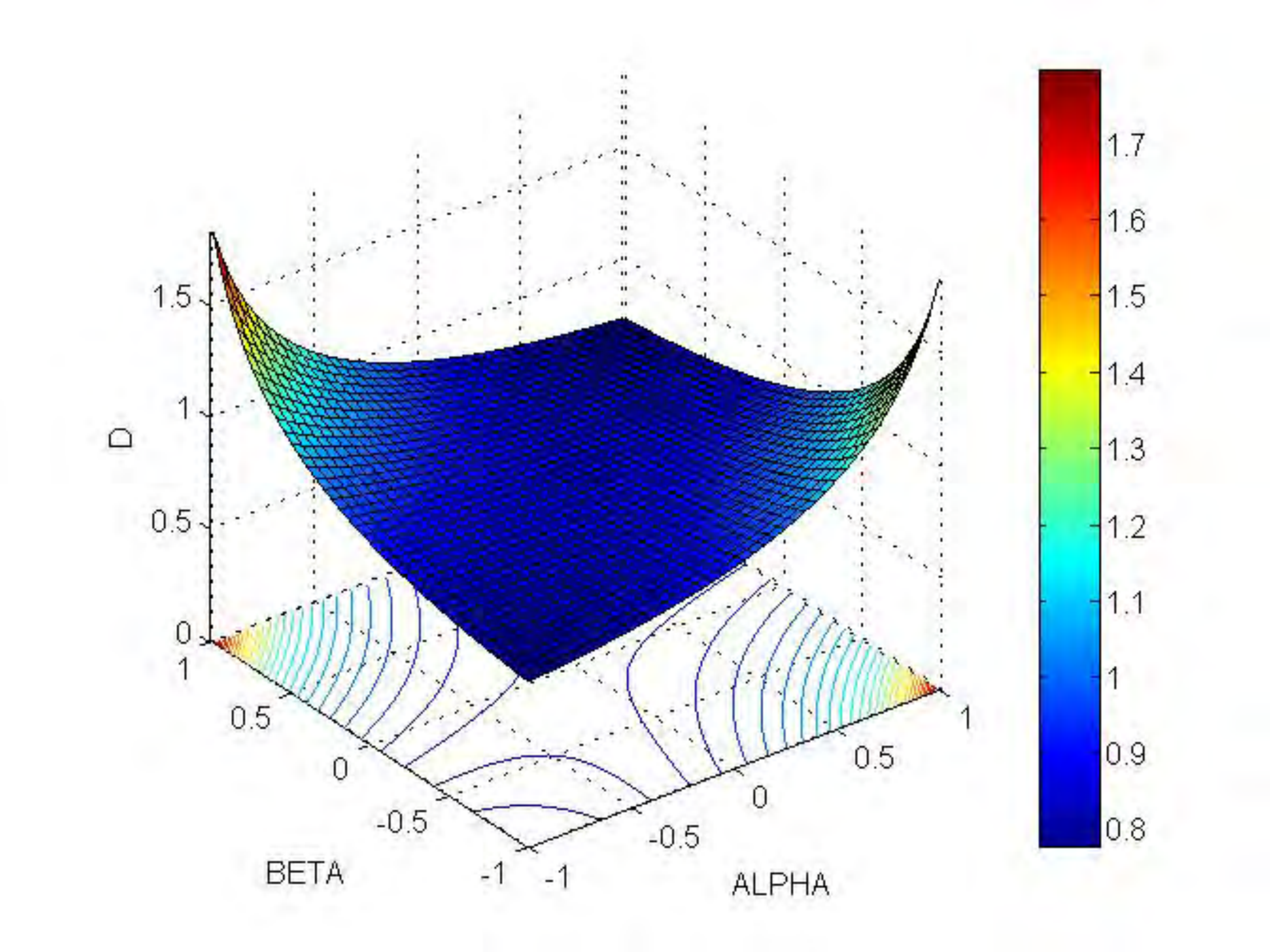}
\hspace{-0.4cm}
(d)
\includegraphics[width=9.3cm]{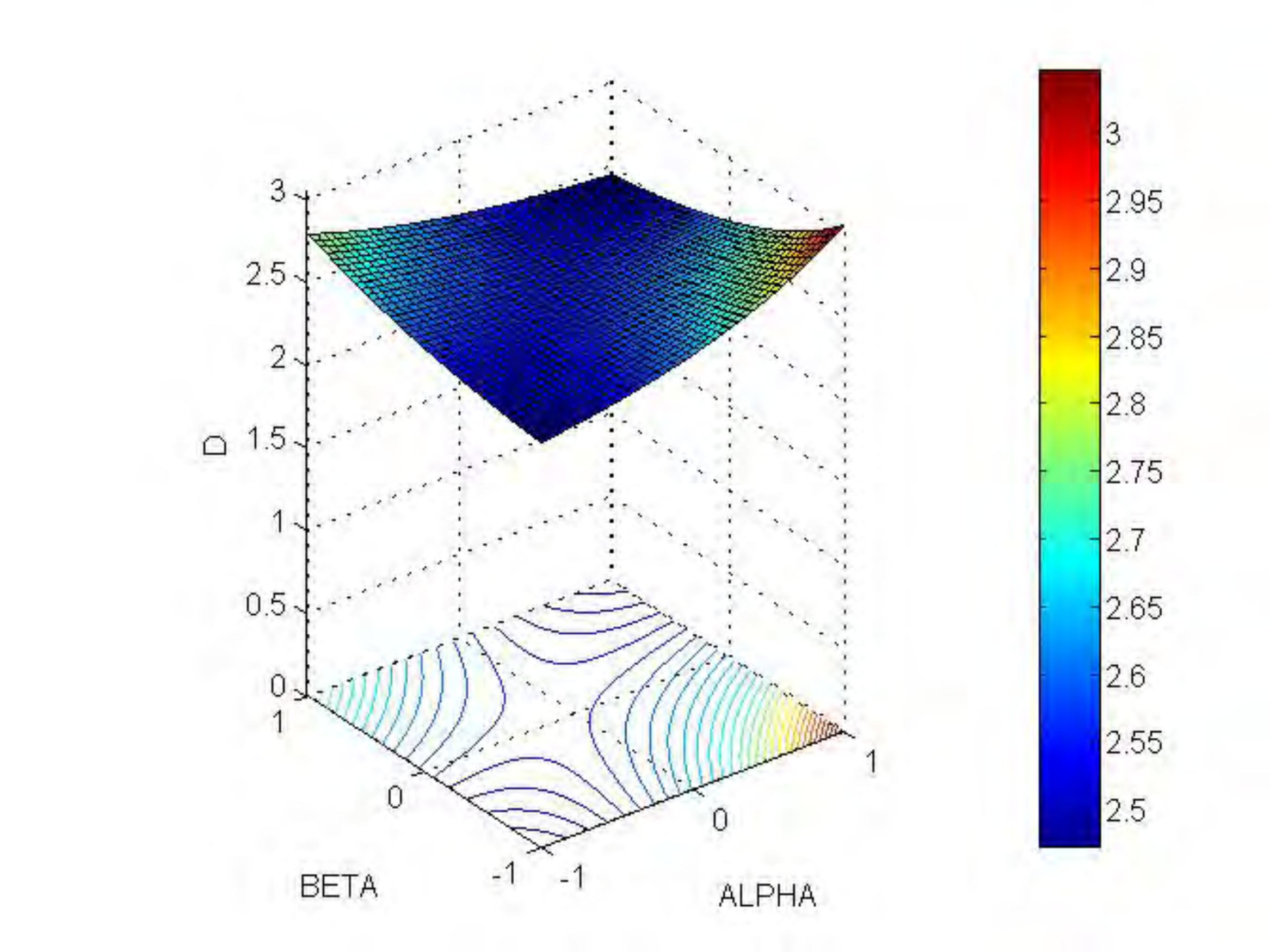}
\caption{2D plots of the AB log-det divergence for different eigenvalues: (a)  $\lambda=0.4$,  (b) $\lambda=2.5$, (c) $\lambda_1=2.5, \lambda_2=0.4$,  (d) for 50 eigenvalues randomly uniformly distributed in the range from 0.5 to 2.}
\label{Fig:2Dplots}
\end{figure}
Fig. \ref{Fig:2Dplots} illustrates typical shapes of the AB log-det divergence for different values of eigenvalues  for a wide range of parameters of $\alpha $ and $\beta$.

In general, the  AB  log-det divergence  is not a metric  distance since  triangular inequality may be not satisfied for some  values of parameters.
Therefore,  we can define optionally a metric distance as a square root of the AB log-det divergence in the special case $\alpha=\beta$ as
\be
d^{(\alpha,\alpha)}_{AB}({\bP}\|{\bQ})=\sqrt{D^{(\alpha,\alpha)}_{AB}({\bP}\|{\bQ})},
\ee
 because $D^{(\alpha,\alpha)}_{AB}({\bP}\|{\bQ})$ is symmetric with respect to $\bP$ and $\bQ$.


As we will show later such defined measures lead to many important divergences and metric distances like the Logdet Zero  divergence, the  AIRM,
squared root of Stein's loss. Moreover, we can generate new divergences, e.g., generalization of Stein's loss, Beta log-det  divergence, or generalized Hilbert metric.

From divergence $D^{(\alpha,\beta)}_{AB}({\bP}\|{\bQ})$, a Riemannian metric and a pair of dually coupled affine connections are introduced in the manifold of positive definite matrices. Let $d\bP$ be a small deviation of $\bP$, which belongs to the tangent space of the manifold at $\bP$. By calculating  $D^{(\alpha,\beta)}_{AB}({\bP + d \bP}\|{\bP})$ and neglecting higher-order terms, we have
\be
 D^{(\alpha,\beta)}_{AB}({\bP+d\bP}\|{\bP})= \frac{1}{2} \tr[d\bP \,\bP^{-1} \,d\bP \,\bP^{-1}].
 \label{RimA}
 \ee
This gives a Riemannian metric which is common for all $(\alpha, \beta)$. Therefore, the Riemannian metric is the same for all AB log-det divergences, although the dual affine connections depend on $\alpha$ and $\beta$. The Riemannian metric is the same as the Fisher information matrix of the manifold of multivariate Gaussian distribution of mean zero and covariance matrix $\bP$.

\begin{figure} [t!]
\begin{center}
\includegraphics[width=17.2cm]{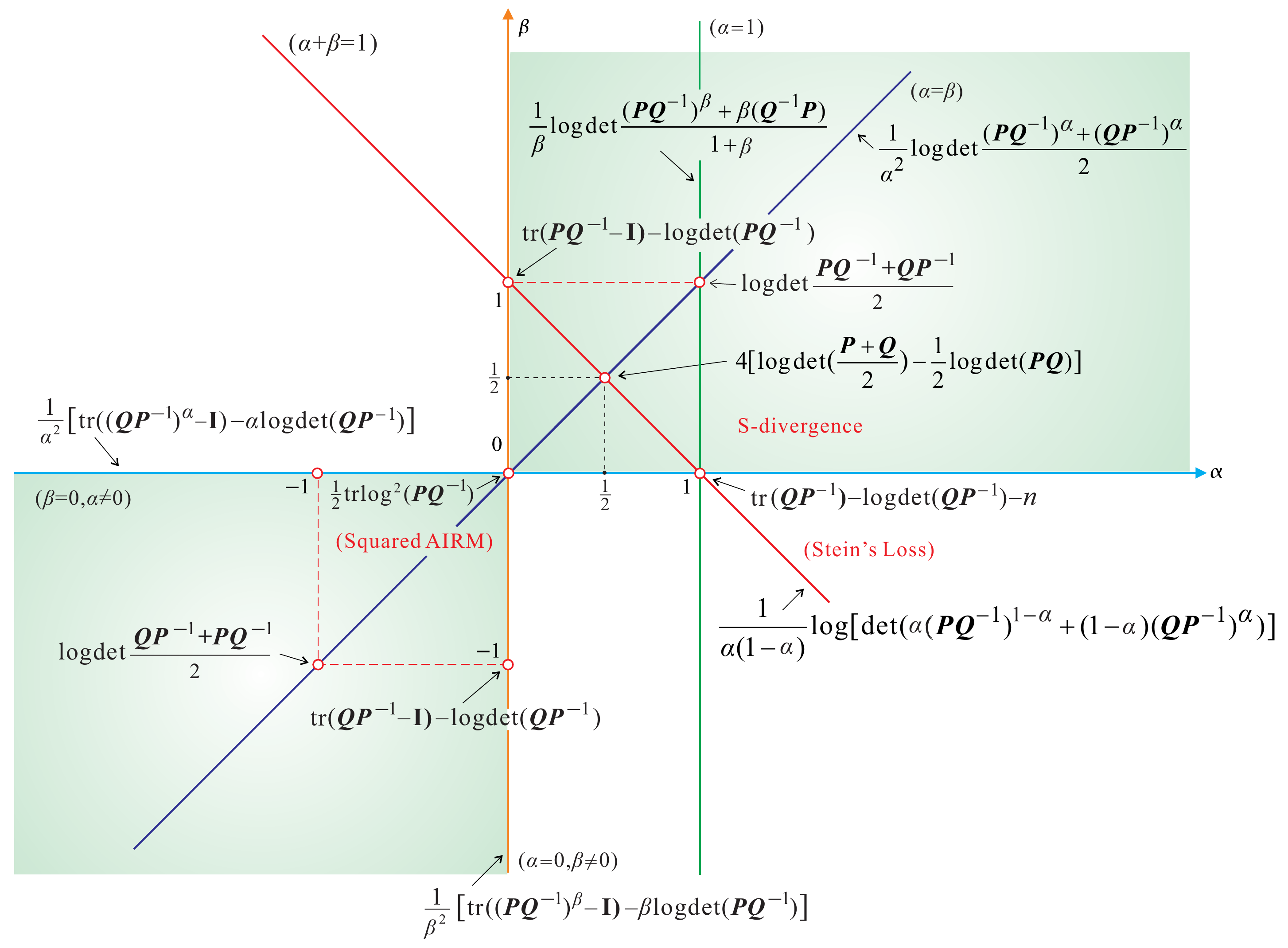}
\caption{Graphical illustration of the fundamental non-symmetric AB log-det divergences.  On the $\alpha$--$\beta$ plane are indicated  important  divergences by points and lines, especially the Stein's loss and its generalization, the AIRM (Riemannian) distance, S-divergence called also Jensen-Bregman LogDet Divergence (JBLD), Alpha log-det divergence $D^{(\alpha)}_A$, and  Beta log-det divergence $D^{(\beta)}_B$.}
\label{Fig1a}
\end{center}
\end{figure}

It is interesting to note that the Riemannian metric or geodesic distance is obtained from
(\ref{defAB1}) for $\alpha=\beta=0$,
%
\be
d_R({\bP}\|{\bQ}) &=& d^{(0,0)}_{AB}({\bP}\|{\bQ}) = \sqrt{D^{(0,0)}_{AB}({\bP}\|{\bQ})} \nonumber \\
&=& \sqrt{\tr \log^2 (\bP \bQ^{-1})} = \sqrt{\tr \log^2 (\bQ \bP^{-1})} \nonumber \\
&=&||\log(\bP \bQ^{-1})||_F= ||\log(\bQ^{-1/2} \bP \bQ^{-1/2})||_F = ||\log(\bP^{-1/2} \bQ \bP^{-1/2})||_F\nonumber \\
&=&\sqrt{\sum_{i=1}^n \log^2 (\lambda_i)},
\label{AIRM1}
\ee
 where $\lambda_i$ are the eigenvalues of the matrix $\bP \bQ^{-1}$.

 This is also known as the Affine Invariant Riemannian metric (AIRM).
AIRM enjoys serval important and useful theoretical properties, and is probably one of the most widely used (dis)similarity measure for SPD (covariance) matrices \cite{Cherian-Sra13},\cite{CherianS14}.

For $\alpha=\beta= 0.5$  (and also for $\alpha=\beta=-0.5$), we obtain the recently defined  and deeply analyzed S-divergence, called also  the JBLD  (Jensen-Bregman LogDet) divergence \cite{Sra-NIPS12},\cite{Sra2014a},\cite{Cherian-Sra13},\cite{CherianS14}:
\be
D_{S}({\bP}\|{\bQ}) &=& D^{(0.5,0.5)}_{AB}({\bP}\|{\bQ})
= 4 \log \displaystyle  \det \left(\frac{1}{2}\left[(\bP \bQ^{-1})^{1/2} +(\bP \bQ^{-1})^{-1/2}\right]\right) \nonumber \\
&=& 4 \log \displaystyle \frac{\det(\bP)^{1/2} \;  \det \left(\displaystyle\frac{(\bP \bQ^{-1})^{1/2} +(\bP \bQ^{-1})^{-1/2}}{2} \right) \; \det(\bQ)^{1/2}}{\det(\bP)^{1/2}\det(\bQ)^{1/2}} \nonumber \\
&=& 4 \log \displaystyle \frac{\det \frac{1}{2} (\bP +\bQ)}{\sqrt{\det(\bP) \det(\bQ)}} \nonumber \\
&=&  4  \left(\log \det \left(\frac{\bP +\bQ}{2} \right) - \frac{1}{2} \log \det(\bP \bQ) \right) =4\sum_{i=1}^n \log \left(\frac{\lambda_i+1}{2 \sqrt \lambda_i}\right).
\label{JBLD1}
\ee
The S-divergence is not metric distance. In order  to make it metric we use square root of it, and we obtain then the LogDet Zero divergence,   called also sometimes the Bhattacharyya distance  \cite{Chebbi2012},\cite{nielsen2013jensen}, \cite{MatrixIG-Nielsen} as
\be
d_{Bh}({\bP}\|{\bQ}) &=& 
\sqrt{D^{(0.5,0.5)}_{AB}({\bP}\|{\bQ})} \nonumber \\
&=&  2 \sqrt{\log \det \left(\frac{\bP +\bQ}{2} \right) - \frac{1}{2} \log \det(\bP \bQ)} \nonumber \\
&=& 2 \sqrt{\log \displaystyle \frac{\det \frac{1}{2} (\bP +\bQ)}{\sqrt{\det(\bP) \det(\bQ)}}} \; .
\label{LDZ}
\ee
Moreover, for $\alpha\neq 0,\;\beta=0$  and for $\alpha =0,\;\beta\neq 0$, we obtain  divergences, which can be considered as  generalizations of Stein's loss (called also Burg matrix divergence or simply LogDet divergence):
\be
 D^{(\alpha,0)}_{AB}({\bP}\|{\bQ}) &=& \displaystyle \frac{1}{\alpha^2} \left[
     \tr \left((\bQ \bP^{-1})^{-\alpha} - \bI\right) + \alpha \log \det (\bQ \bP^{-1}) \right], \;\;\; \alpha\neq0
 \label{genSteinloss1}
 \ee
\be
 D^{(0,\beta)}_{AB}({\bP}\|{\bQ}) =   \displaystyle  \frac{1}{\beta^2} \left[
     \tr \left((\bP \bQ^{-1})^{\beta} - \bI \right) - \beta \log \det (\bP \bQ^{-1}) \right],  \;\; \; \beta \neq 0.
  \label{genSteinloss2}
    \ee
 The  divergences (\ref{genSteinloss1}) and (\ref{genSteinloss2}) can be simplified to the standard Stein's loss  for $\alpha=1$ and $\beta=1$, respectively.

One important potential application of the AB  log-det divergence is to generate efficient  conditionally positive definite kernels, which can be found wide applications in classification and clustering. It seems that for a specific set of parameters the AB log-det  divergence divergences admit a Hilbert space embedding in the form of a Radial Basis Function (RBF) kernel \cite{Kulis}. More specifically, it can be shown that AB  log-det kernel can be
defined as
\be
K^{(\alpha,\beta)}_{AB}({\bP}\|{\bQ})&=&
\exp\left( -\gamma D^{(\alpha,\beta)}_{AB}({\bP}\|{\bQ})\right) \nonumber \\
&=& \left( \det
    \displaystyle \frac{\alpha (\bP \bQ^{-1})^{\beta} + \beta (\bQ \bP^{-1})^{\alpha}}
        {\alpha+\beta} \right)^{-\frac{\gamma}{\alpha \; \beta}}
\ee
where $\gamma >0$ and  $ \alpha,\beta>0 \;\;  \mbox{or} \;\; \alpha,\beta<0$, which some selected values of $\gamma$ parameters  is
positive definite. However, the topic of  kernel properties and their applications is out of the scope of this review paper.

\section{Special Cases of the AB Log-Det Divergence}

We shall now illustrate that a suitable choice of the $(\alpha,\beta)$ parameters simplifies the AB log-det divergence into some known divergences, including the Alpha- and Beta- log-det divergences \cite{Chebbi2012}, \cite{CherianSBPLAL2011},\cite{nielsen2013jensen},\cite{Cich-Amari-entropy}.

When $\alpha+\beta=1$ the AB log-det  divergence reduces to the Alpha-log-det divergence \cite{Chebbi2012}
\begin{eqnarray}  \label{AlphaLD1}
    D^{(\alpha,1-\alpha)}_{AB} ({\bP} \|  {\bQ})
    &=&
    D^{(\alpha)}_{A}\left({\bP}\|{\bQ}\right)   \\&\doteq&
    \left\{
    \begin{tabular}{ll}
    $\displaystyle \frac{1}{\alpha(1- \alpha)} \log \det \left[ \alpha (\bP \bQ^{-1})^{1-\alpha} +(1-\alpha)(\bQ \bP^{-1})^{\alpha} \right] = $   & 
   \nonumber  \\
    $\displaystyle \frac{1}{\alpha(1- \alpha)} \log \displaystyle \frac{\det \left(\alpha \bP +(1-\alpha) \bQ \right)}{\det \left(\bP^{\alpha} \; \bQ^{1-\alpha} \right)}= $ &  \nonumber  \\
    $\displaystyle  \frac{1}{\alpha (1-\alpha)}  \sum_{i=1}^n \log \left(\frac
     {\alpha (\lambda_i-1)  + 1}
        {\lambda_i^{\alpha}}\right) $   &\hspace{-0.5cm} $\ \text{for}\ \;\; 0<\alpha<1 $\\
    $ \tr(\bQ \bP^{-1}) -\log \det(\bQ \bP^{-1}) -n=\displaystyle\sum_{i=1}^n \left(\lambda^{-1}_i+\log(\lambda_i)\right) -n $   &$\ \text{for}\ \;\; \alpha=1 $,
    \\
    $ \tr( \bP \bQ^{-1}) -\log \det( \bP \bQ^{-1}) -n = \displaystyle\sum_{i=1}^n \left(\lambda_i-\log(\lambda_i)\right) -n$     &$\ \text{for}\ \;\; \alpha= 0 $
    \end{tabular}
    \right.
\end{eqnarray}
On the other hand, when $\alpha=1$, and $\beta \geq 0$ the AB log-det  divergence reduces to the Beta- log-det divergence
\begin{eqnarray}
    D^{(1,\beta)}_{AB} \left({\bP}   \|  {\bQ}\right)  &=&  D^{(\beta)}_{B}({\bP}\|{\bQ})
    \\&\doteq&
    \left\{
    \begin{tabular}{ll}
$\displaystyle \frac{1}{\beta} \log  \det \displaystyle\frac{(\bP \bQ^{-1})^{\beta} + \beta \; ( \bQ \bP^{-1})}{1+\beta}=\displaystyle  \frac{1}{ \beta}  \sum_{i=1}^n \log \left(\frac
     { \lambda_i^{\beta}  + \beta \lambda_i^{-1}}
        {1+\beta}\right) $   &$\ \text{for}\ \;\; \beta >0 $,
   \\
    $ \tr(\bQ \bP^{-1}-\bI) -\log \det(\bQ \bP^{-1})=\displaystyle\sum_{i=1}^n \left(\lambda^{-1}_i+\log(\lambda_i)\right) -n $     &$\ \text{for}\ \;\; \beta= 0 $, \nonumber\\
     $ \log \displaystyle \frac{\det(\bP \bQ^{-1})}{\det(\bI +\log(\bP\bQ^{-1}))}= \displaystyle \sum_{i=1}^n  \log \displaystyle \frac{\lambda_i}{1+\log(\lambda_i)}$  & \hspace{-1.99cm}$\ \text{for}\ \;\; \beta= -1,\; \lambda_i > e^{-1} \forall i $
    \end{tabular}
    \right.
    \label{BetaLD1}
\end{eqnarray}
It should be noted that $\det(\bI +\log(\bP\bQ^{-1})= \prod_{i=1}^n [1 +\log (\lambda_i)]$ and the Beta  log-det divergence is well defined for $\beta=-1$ if all eigenvalues are larger than $\lambda_i > e^{-1} \approx 0.367$ ($e\approx 2.72$).

It is interesting to note that the Beta log-det divergence  for $\beta \rightarrow \infty$ leads to a new (robust in respect to noise) divergence expressed as{\footnote{ This can be easily shown by applying L'H\^opital's formula.}}
\be
\lim_{\beta \rightarrow \infty} D^{(\beta)}_{B}({\bP}\|{\bQ})
    =  D^{(\infty)}_{B}({\bP}\|{\bQ})= \log (\prod_{i=1}^k \lambda_i) \;\; \mbox{for all} \;\; \lambda_i \geq 1.
\ee
Assuming that the set $\Omega=\{i: \lambda_i > 1\}$,  gathers the indices of those eigenvalues greater than one, we can more formally express such divergence as
\be
 D^{(\infty)}_{B}({\bP}\|{\bQ})
    &=&
    \left\{
    \begin{tabular}{ll}
	$\log (\prod_{i\in \Omega} \lambda_i)$ &	 for $\Omega\neq \phi$\\
    $0$ & for\ $\Omega= \phi$.
    \end{tabular}
    \right. 	
\ee
%
The Alpha-log-det divergence gives  the standard Stein's losses (Burg matrix divergences) for $\alpha=1$ and $\alpha=0$  and
the Beta-log-det divergence  is also the Stein's loss  for  $\beta=0$.



Another important class of divergences is Power log-det  divergence for any $\alpha=\beta \in \Real$
\begin{eqnarray}
    D^{(\alpha,\alpha)}_{AB} \left({\bP}   \|  {\bQ}\right)  &=&  D^{(\alpha)}_{P}({\bP}\|{\bQ})
    \\&\doteq&
    \left\{
    \begin{tabular}{ll}
$\displaystyle \frac{1}{\alpha^2} \log  \det \displaystyle \frac{(\bP \bQ^{-1})^{\alpha} +  (\bP \bQ^{-1})^{-\alpha}}{2} = \displaystyle \frac{1}{\alpha^2} \sum_{i=1}^n \log \frac{\lambda_i^{\alpha}+\lambda_i^{-\alpha}}{2}$   &$\ \text{for}\ \;\; \alpha \neq 0 $, \nonumber \\
   \\
    $ \displaystyle \frac{1}{2} \tr \log^2\det(\bP \bQ^{-1}) = \displaystyle \frac{1}{2} \tr \log^2\det(\bQ \bP^{-1}) = \displaystyle\frac{1}{2} \sum_{i=1}^n \log^2 (\lambda_i) $     &$\ \text{for}\ \;\; \alpha = 0 $.
    \end{tabular}
    \right.
    \label{PowerLD1}
\end{eqnarray}

\section{Fundamental Properties of the AB Log-Det Divergence}

The AB log-det divergence has several important and useful theoretical properties for any SPD matrices

\begin{enumerate}

\item Nonnegativity
\be
D^{(\alpha,\beta)}_{AB}({\bP}\|{\bQ}) \geq 0, \;\;\mbox{for} \;\; \alpha \geq 0 \;\; \mbox{and} \; \beta \geq0\;\; \mbox{or} \;\; \alpha \leq 0 \;\; \mbox{and} \; \beta \leq 0.
\ee

\item Definiteness (see Theorem 1 and 2)
\be
D^{(\alpha,\beta)}_{AB}({\bP}\|{\bQ})= 0 \;\; \mbox{iff} \;\; \bP=\bQ.
\ee

\item Continuity and smoothness of the  $D^{(\alpha,\beta)}_{AB}({\bP}\|{\bQ})$  as function of  parameters
$\alpha$ and $\beta$ in the whole space including singular values $\alpha \neq 0$, $\beta \neq 0$ and
$\alpha =-\beta$ (see Fig. \ref{Fig:2Dplots}).

\item The divergence can be explicitly expressed by eigenvalues of the matrix $\bQ^{-1}\bP$
\be
 D^{(\alpha,\beta)}_{AB}({\bP }\|{\bQ }) =  D^{(\alpha,\beta)}_{AB}({\bQ^{-1} \bP}\|{\bI}) =  D^{(\alpha,\beta)}_{AB}({\mbi \Lambda}\|{\bI}),
  \ee
where $\mbi \Lambda= \diag\{\lambda_1,\lambda_2, \ldots, \lambda_n\}$.

Proof:
From the definition of the divergence it is evident that $D^{(\alpha,\beta)}_{AB}({\bP }\|{\bQ })=D^{(\alpha,\beta)}_{AB}(\bP \bQ^{-1}\|{\bI })$.
Then, taking into account the eigenvalue decomposition  $\bP \bQ^{-1} = \bV \mbi \Lambda \; \bV^{-1}$, we can write
\be
 D^{(\alpha,\beta)}_{AB}({\bP}\|{\bQ})
   &=&
    \frac{1}{\alpha \beta} \log \det
     \frac{\alpha  \; \bV \mbi \Lambda^{\beta} \; \bV^{-1} + \beta \; \bV \mbi \Lambda^{-\alpha} \; \bV^{-1}}
        {\alpha+\beta} \nonumber \\
         &=&  \frac{1}{\alpha \beta} \log \left[ \det \bV \;
     \det \frac{\alpha \mbi \Lambda^{\beta}  + \beta \mbi \Lambda^{-\alpha}}
        {\alpha+\beta} \; \det \bV^{-1}\right] \nonumber \\
        &=&  \frac{1}{\alpha \beta} \log \det
     \frac{\alpha \; \mbi \Lambda^{\beta}  + \beta \; \mbi \Lambda^{-\alpha}}
        {\alpha+\beta}\\
      &=&D^{(\alpha,\beta)}_{AB}(\mbi \Lambda \|{\bI })
\ee
  \item Scaling invariance
  \be
  D^{(\alpha,\beta)}_{AB}({c \bP }\|{c \bQ }) =  D^{(\alpha,\beta)}_{AB}({\bP}\|{\bQ})
  \ee
for any $c>0$, or more general
\be
  D^{(\alpha,\beta)}_{AB}({\bP \bC}\|{\bQ \bC}) = D^{(\alpha,\beta)}_{AB}({\bP}\|{\bQ})
  \ee
for any nonsingular matrix $\bC \in \Real^{n \times n}$.

Proof:
\be
  D^{(\alpha,\beta)}_{AB}({\bP \bC}\|{\bQ \bC})
  &=& D^{(\alpha,\beta)}_{AB}({\bP \bC}({\bQ \bC})^{-1}\|\bI)\\
  &=& D^{(\alpha,\beta)}_{AB}({\bP}\bQ^{-1}\|\bI)\\
  &=& D^{(\alpha,\beta)}_{AB}({\bP}\|{\bQ})\; .
\ee

\item For a given $\alpha,\beta$ parameters and a non-zero scaling scalar $\omega \neq0$,
 \be
  D^{(\omega \; \alpha, \; \omega \;\beta)}_{AB}({ \bP }\|{\bQ }) =
  \frac{1}{\omega^2}  D^{(\alpha,\beta)}_{AB}({\bP^{\; \omega}}\|{\bQ^{\;\omega}})\; .
  \ee

Proof:
From the definition of the divergence we can write
\be
 D^{(\omega \; \alpha, \; \omega \;\beta)}_{AB}({ \bP }\|{\bQ })
        &=&  \frac{1}{(\omega \alpha)(\omega \beta)} \log \det
     \frac{\omega\alpha \; \mbi \Lambda^{\omega\beta}  + \omega\beta \; \mbi \Lambda^{-\omega\alpha}}
        {(\omega\alpha+\omega\beta)}\\
       &=&  \frac{1}{\omega^2}\frac{1}{ \alpha \beta} \log \det
     \frac{\alpha \; (\mbi \Lambda^{\omega})^{\beta}  + \beta \; (\mbi \Lambda^{\omega})^{-\alpha}}
        {(\alpha+\beta)}\\
      &=&\frac{1}{\omega^2}  D^{(\alpha,\beta)}_{AB}({\bP^{\; \omega}}\|{\bQ^{\;\omega}})\; .
\ee

 Hence, we can obtain important inequality
  \be
 D^{(\alpha,\beta)}_{AB}({\bP^{\; \omega}}\|{\bQ^{\;\omega}}) &\leq&  D^{(\omega \; \alpha, \; \omega \;\beta)}_{AB}({ \bP }\|{\bQ })
  \ee
 for $|\omega| \leq 1 $. 

 \item Dual--invariance under inversion (for $\omega=-1$)
   \be
  D^{(-\alpha,-\beta)}_{AB}({\bP }\|{ \bQ }) =  D^{(\alpha,\beta)}_{AB}({\bP^{-1}}\|{\bQ^{-1}})\; .
  \ee

\item Dual symmetry
  \be
  D^{(\alpha,\beta)}_{AB}({\bP }\|{ \bQ }) =  D^{(\beta,\alpha)}_{AB}({\bQ}\|{\bP})\; .
  \ee

 \item Affine invariance (invariance under linear transformations)
 \be
  D^{(\alpha,\beta)}_{AB}({\bA \bP \bB} \|{ \bA \bQ \bB}) &=&  D^{(\alpha,\beta)}_{AB}({\bP}\|{\bQ})
  \ee
for any nonsingular matrices $\bA \in \Real^{n \times n}$ and  $\bB \in \Real^{n \times n}$,

Proof:
\be
 D^{(\alpha,\beta)}_{AB}({\bA \bP \bB} \|{ \bA \bQ \bB})
   &=&
    \frac{1}{\alpha \beta} \log \det
     \frac{\alpha  \; (({\bA \bP \bB})({\bA \bQ \bB})^{-1})^{\beta}
     + \beta \; (({\bA \bP \bB})({\bA \bQ \bB})^{-1})^{-\alpha}}
        {\alpha+\beta} \nonumber \\
 &=&
    \frac{1}{\alpha \beta} \log \det
     \frac{\alpha  \; (\bA (\bP \bQ^{-1}) \bA^{-1})^{\beta}
     + \beta \; (\bA (\bP \bQ^{-1}) \bA^{-1})^{-\alpha}}
        {\alpha+\beta} \nonumber \\
         &=&  \frac{1}{\alpha \beta} \log \left[ \det (\bA\bV) \;
     \det \frac{\alpha \mbi \Lambda^{\beta}  + \beta \mbi \Lambda^{-\alpha}}
        {\alpha+\beta} \; \det (\bA\bV)^{-1}\right] \nonumber \\
        &=&  \frac{1}{\alpha \beta} \log \det
     \frac{\alpha \; \mbi \Lambda^{\beta}  + \beta \; \mbi \Lambda^{-\alpha}}
        {\alpha+\beta}\\
      &=&D^{(\alpha,\beta)}_{AB}({\bP}\|{\bQ})\; .
\ee

\item Scaling invariance under Kronecker product
\be
  D^{(\alpha,\alpha)}_{AB}({\bA \otimes \bP} \|{ \bA  \otimes \bQ }) =  n  D^{(\alpha,\alpha)}_{AB}({\bP}\|{\bQ})\; .
  \ee

Proof:
\be
  D^{(\alpha,\alpha)}_{AB}({\bA \otimes \bP} \|{ \bA  \otimes \bQ })
   &=&D^{(\alpha,\beta)}_{AB}(({\bA \otimes \bP})  ({ \bA  \otimes \bQ })^{-1}\|{\bI })\\
   &=&D^{(\alpha,\beta)}_{AB}(({\bA \bA^{-1}})\otimes ({ \bP \bQ }^{-1})\|{\bI })\\
   &=&
    \frac{1}{\alpha \beta} \log \det
    	\left[
      \bI \otimes \frac{\alpha  \; (\bP \bQ^{-1})^{\beta}
     + \beta \; (\bP \bQ^{-1})^{-\alpha}}
        {\alpha+\beta}
    \right]     \nonumber \\
         &=& \frac{1}{\alpha \beta} \log \det
      \left[
      \frac{\alpha  \; (\bP \bQ^{-1})^{\beta}
     + \beta \; (\bP \bQ^{-1})^{-\alpha}}
        {\alpha+\beta}
         \right]^n
         \nonumber \\
      &=& n\; D^{(\alpha,\beta)}_{AB}({\bP}\|{\bQ})\; .
\ee

  \item Triangle Inequality -- Metric Distance Condition 
\be
\sqrt{D^{(\alpha,\alpha)}_{AB}({\bP}\|{\bQ})} \leq \sqrt{D^{(\alpha,\alpha)}_{AB}({\bP}\|{\bZ})} +\sqrt{D^{(\alpha,\alpha)}_{AB}({\bZ}\|{\bQ})}\; .
\ee
Proof:
  On the one hand, for $\alpha\neq 0$, we can prove the metric condition with the help of the Bhattacharryya distance
\be
 d_{Bh}({ \bP }\|{\bQ })
  =\sqrt{D^{( 0.5, \; 0.5)}_{AB}({ \bP }\|{\bQ })}\\
  = 2 \sqrt{\log \displaystyle \frac{\det \frac{1}{2} (\bP +\bQ)}{\sqrt{\det(\bP) \det(\bQ)}}} \; .
\ee
By defining $\omega=2\alpha\neq 0$ and using the property
\be
 \sqrt{D^{( \alpha, \; \alpha)}_{AB}({ \bP }\|{\bQ })}
  &=&
 \sqrt{D^{(\omega\; 0.5, \; \omega\; 0.5)}_{AB}({ \bP }\|{\bQ })}  \\
      &=&\sqrt{\frac{1}{\omega^2}  D^{(0.5,0.5)}_{AB}({\bP^{\; \omega}}\|{\bQ^{\;\omega}})}\\
      &=&\frac{1}{2	|\alpha|}  \sqrt{D^{(0.5,0.5)}_{AB}({\bP^{\; 2\alpha}}\|{\bQ^{\; 2\alpha}})}\\
      &=&\frac{1}{2	|\alpha|}  d_{Bh}({\bP^{\; 2\alpha}}\|{\bQ^{\; 2\alpha}})\; ,
\ee
the metric condition can be easily verified. For instance, in order to check the triangle inequality we can observe that
\be
\sqrt{D^{(\alpha,\alpha)}_{AB}({\bP}\|{\bQ})}
&=&\frac{1}{2	|\alpha|} d_{Bh}({\bP^{\; 2\alpha}}\|{\bQ^{\; 2\alpha}})\\
&\leq&
\frac{1}{2	|\alpha|}d_{Bh}({\bP}^{\; 2\alpha}\|{\bZ}^{\; 2\alpha}) +d_{Bh}({\bZ}^{\; 2\alpha}\|{\bQ}^{\; 2\alpha})
\\
&=&
\sqrt{D^{(\alpha,\alpha)}_{AB}({\bP}\|{\bZ})} +\sqrt{D^{(\alpha,\alpha)}_{AB}({\bZ}\|{\bQ})}\; .
\ee
On the other hand, $\sqrt{D^{(\alpha,\alpha)}_{AB}({\bP}\|{\bQ})}$ for $\alpha\rightarrow0$ converges to the Riemannian metric
\be
    \sqrt{D^{(0,0)}_{AB}({\bP}\|{\bQ})}
	&=& \lim_{\alpha\rightarrow 0}\sqrt{D^{(\alpha,\alpha)}_{AB}({\bP}\|{\bQ})} \\
	&=& \|\log(\bQ^{-1/2} \bP \bQ^{-1/2})\|_F \\
	&=& d_R({\bP}\|{\bQ})\; ,	
\ee
which concludes the proof of the metric condition of $\sqrt{D^{(\alpha,\alpha)}_{AB}({\bP}\|{\bQ})}$ for any $\alpha\in \mathbb{R}$.

  \end{enumerate}

\section{Symmetrized  AB Log-Det Divergences}

The basic AB log-det  divergence is asymmetric, that is,
$D^{(\alpha,\beta)}_{AB}(\bP\, || \, \bQ) \neq D^{(\alpha,\beta)}_{AB}(\bQ \,|| \,\bP)$, except the spacial case of $\alpha=\beta$).

Generally, there are several ways to symmetrize a divergence, for example:
Type-1
 \be
D^{(\alpha,\beta)}_{ABS1}(\bP\, || \, \bQ)=\frac{1}{2} \left[D^{(\alpha,\beta)}_{AB}(\bP\, || \, \bQ) + D^{(\alpha,\beta)}_{AB}(\bQ \,|| \,\bP) \right]
\ee
and Type-2 based on Jensen-Shannon symmetrization (which seems to be too complex for log-det divergences)
 \be
D^{(\alpha,\beta)}_{ABS2}(\bP\, || \, \bQ)=\frac{1}{2} \left[D^{(\alpha,\beta)}_{AB} \left(\bP\, || \, \frac{\bP+\bQ}{2} \right) + D^{(\alpha,\beta)}_{AB} \left(\bQ \,|| \,\frac{\bP+\bQ}{2}\right) \right].
\ee

The symmetric AB log-det divergence (Type-1) can be  defined as 
\be
    \label{ABdef-full2}
    D^{(\alpha,\beta)}_{ABS1}({\bP}\|{\bQ}) &=&
    \left\{
    \begin{tabular}{ll}
    $\displaystyle  \frac{1}{2 \alpha \beta} (\log \det
     \frac{\alpha (\bP \bQ^{-1})^{\beta} + \beta (\bQ \bP^{-1})^{\alpha}}
        {\alpha+\beta} + $ & $$ \\ \\
        $+ \log \det \displaystyle
     \frac{\alpha (\bQ \bP^{-1})^{\beta} + \beta (\bP \bQ^{-1})^{\alpha}}
        {\alpha+\beta}) $&  $\mbox{for} \;\; \alpha,\beta>0 \;\;  \mbox{or} \;\; \alpha,\beta<0$
        \\
    \\
    $ \displaystyle \frac{1}{2 \alpha^2} \left[
     \tr \left((\bP \bQ^{-1})^{\alpha} + (\bQ \bP^{-1})^{\alpha} - 2 \bI\right)  \right]
         $   &$\ \text{for}\ \;\; \alpha \neq 0, \; \beta= 0 $ \\
    \\
    $\displaystyle  \frac{1}{2 \beta^2} \left[
     \tr \left((\bP \bQ^{-1})^{\beta} + (\bQ \bP^{-1})^{\beta} - 2\bI \right)  \right] $   &$\
      \text{for}\ \;\; \alpha= 0, \; \beta\neq 0 $ \\
       \\
    $\displaystyle \frac{1}{2\alpha^2}
      \tr\log \displaystyle (\bI -\log^2(\bP\bQ^{-1})^\alpha)^{-1}$
      &$\
      \text{for}\ \;\; \alpha=-\beta\neq 0 $ \\\\
    $\displaystyle \frac{1}{2} \tr \log^2 (\bP \bQ^{-1}) = \frac{1}{2} ||\log(\bQ^{-1/2} \bP \bQ^{-1/2})||^2_F$ & $\ \text{for}\ \;\; \alpha, \; \beta= 0 $.
    \end{tabular}
    \right.
\ee
or equivalently expressed by eigenvalues of the matrix $\bP \bQ^{-1}$:
\be
    \label{ABdef-full-lambda}
    D^{(\alpha,\beta)}_{ABS1}({\bP}\|{\bQ}) &=&
    \left\{
    \begin{tabular}{ll}
    $\displaystyle  \frac{1}{2\alpha \beta}  \sum_{i=1}^n \log \left(1+\frac
     {\alpha\beta}{(\alpha+\beta)^2} (\lambda_i^{\alpha+\beta}  + \lambda_i^{-(\alpha+\beta)}-2)
     \right)$   &$\mbox{for} \; \alpha,\beta>0 \;  \mbox{or} \; \alpha,\beta<0$ \\
    \\
    $ \displaystyle \frac{1}{2 \alpha^2} \left[ \sum_{i=1}^n \left( \lambda_i^{\alpha} + \lambda_i^{-\alpha} \right) -2n \right]=\displaystyle \frac{1}{2 \alpha^2}  \sum_{i=1}^n
    \frac{(\lambda_i^{\alpha}-1)^2}{\lambda_i^{\alpha}}
         $   &$\ \text{for}\ \;\; \alpha \neq 0, \; \beta= 0 $ \\
    \\
    $\displaystyle  \frac{1}{2 \beta^2} \left[ \sum_{i=1}^n \left(\lambda_i^{\beta} + \lambda_i^{-\beta}\right) -2 n \right]=\displaystyle \frac{1}{2 \beta^2}  \sum_{i=1}^n
    \frac{(\lambda_i^{\beta}-1)^2}{\lambda_i^{\beta}} $   &$\
      \text{for}\ \;\; \alpha= 0, \; \beta\neq 0 $ \\
       \\
         $\displaystyle  \frac{1}{2 \alpha^2} \sum_{i=1}^n \log\frac{1}{1-\log^2( \lambda_i^{\alpha})}$
      &$\
      \text{for}\ \;\; \alpha=-\beta\neq 0 $ \\\\
    $\displaystyle  \frac{1}{2}  \sum_{i=1}^n \log^2 (\lambda_i) $ & $\ \text{for}\ \;\; \alpha, \; \beta= 0 $.
    \end{tabular}
    \right.
\ee
\begin{figure} [t!]
\begin{center}
\includegraphics[width=17.0cm]{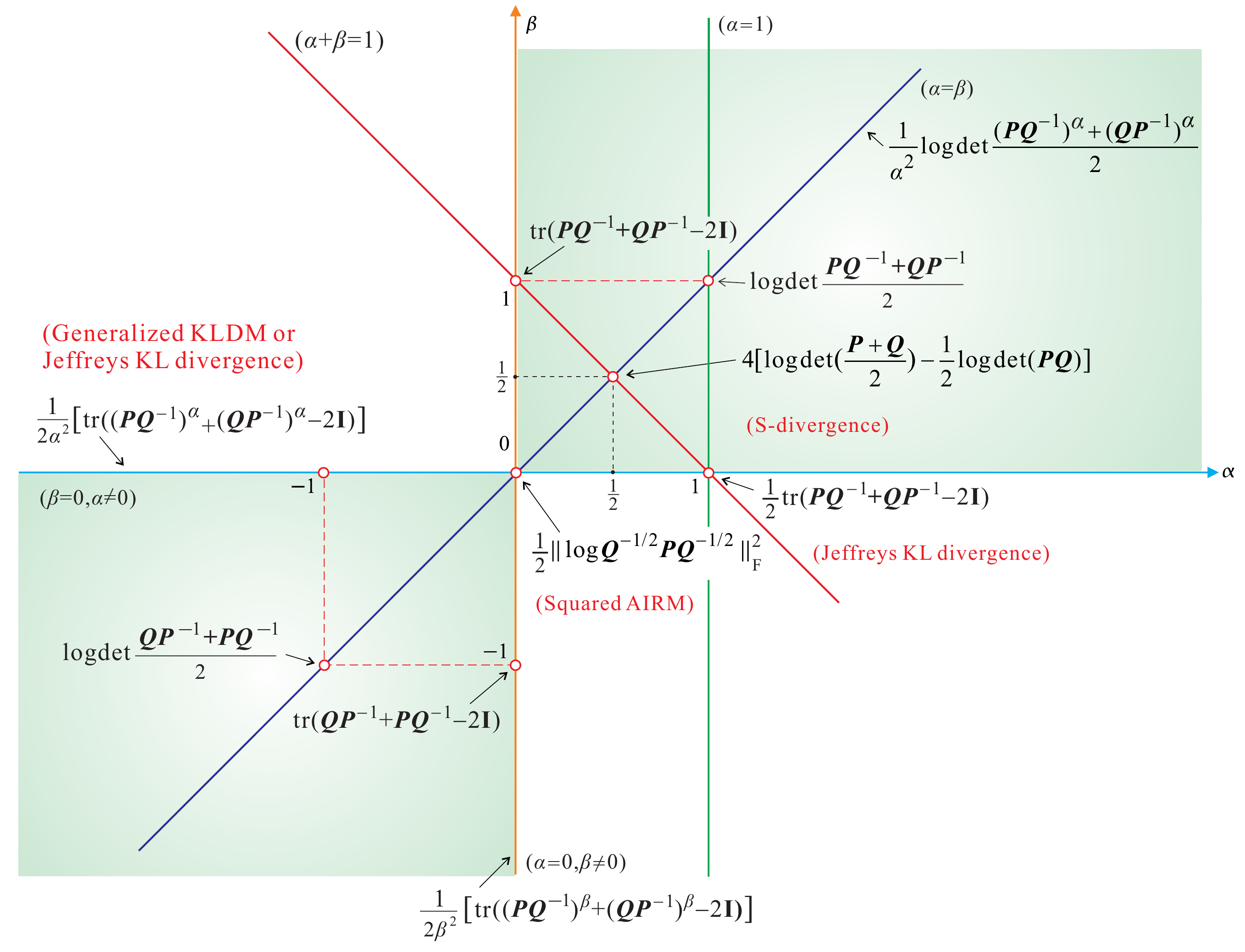}
\caption{Graphical illustration of the fundamental symmetric AB log-det divergences.  On the alpha-beta plane are indicated as special important cases  particular divergences by points, especially  Jeffreys KL divergence, called also KLDM (KL Divergence Metric) or symmetric Stein's loss and its generalization, S-divergence or  JBLD-divergence, and Power log-det divergence.}
\label{ABLD2}
\end{center}
\end{figure}
As special cases, we obtain several well-known symmetric log-det divergences (see Fig. \ref{ABLD2}), for example :\\

(1) For $\alpha=\beta= \pm 0.5$, we obtain the S-divergence or the JBLD divergence (\ref{JBLD1})

(2) For $\alpha=\beta=0$, we have the square of the  AIRM (Riemannian metric) (\ref{AIRM1}).

(2) For $\alpha =0$ and $\beta= \pm 1$ and $\beta=0$ and $\alpha= \pm 1$, we obtain the KLDM (symmetrized KL Density Metric), called also the symmetric Stein's loss or Jeffreys KL divergence:
\be
D_{JKL}(\bP \| \bQ) &=& \displaystyle \frac{1}{2} \tr\left(\bP \bQ^{-1} + \bQ \bP^{-1} -2 \; \bI \right) \nonumber \\
&=& \displaystyle \frac{1}{2} \tr\left(\bP \bQ^{-1} + \bQ \bP^{-1}\right) -n \nonumber \\
&=& \displaystyle \frac{1}{2} \sum_{i=1}^n \left( \sqrt{\lambda_i} - \frac{1}{\sqrt{\lambda_i}}\right)^2.
\ee

\section{Modifications and Generalizations of AB Log-Det Divergences, Gamma Matrix Divergences}

The divergence (\ref{defAB1}) discussed in previous sections can be extended or modified in several ways.

First of all,  we can define alternative AB log-det divergence as follows
\begin{eqnarray}
 \label{defAB5}
 \widetilde{D}^{(\alpha,\beta)}_{AB}({\bP}\|{\bQ})
  & = &
    \frac{1}{\alpha \beta} \log
    \displaystyle
     \frac{\det \left(\displaystyle \frac{\alpha \;(\bP)^{\alpha+\beta} + \beta \; (\bQ)^{\alpha+\beta}}{\alpha +\beta}\right)}
        {\det (\bP)^{\alpha} \det (\bQ)^{\beta}} \\ 
         &&\ \text{for}\ \alpha \neq 0, \;\; \beta\neq 0, \;\;\; \alpha+\beta\neq 0, \;\; \alpha>0 , \;\beta>0 \nonumber
\end{eqnarray}
%
It can be shown that for $\alpha+\beta=1$ (i.e., for Alpha log-det divergence - see Eq. (\ref{defAB1})):
\be
 \widetilde{D}^{(\alpha,\beta)}_{AB}({\bP}\|{\bQ})= {D}^{(\alpha,\beta)}_{AB}({\bP}\|{\bQ})
\ee
However,  they are not equivalent in more general cases.
In fact, it is easy to show that the divergence (\ref{defAB5}) can be expressed as a scaled and transformed Alpha log-det  divergence of the form (see (\ref{AlphaLD1}))
%
\be
 \widetilde{D}^{(\alpha,\beta)}_{AB}({\bP}\|{\bQ})= (\alpha+\beta)^2 {D}^{(\frac{\alpha}{\alpha+\beta})}_{A}({\bP}^{\frac{\alpha}{\alpha+\beta}} \|{\bQ}^{\frac{\alpha}{\alpha+\beta}}),
\ee
so (\ref{defAB5}) is  less general than  (\ref{defAB1}), since it does not cover Power and Beta log-det divergences.

It is interesting to note  that positive eigenvalues of the matrix $\bP \bQ^{-1}$
play similar role to ratios $(p_i/q_i)$ and   $(q_i/p_i)$ used in the wide class of standard discrete divergences, see for example, \cite{Cich-Cruces-Am},\cite{Cich-Amari-entropy}, so we can apply such divergences  to formulate modified log-det divergence as functions of eigenvalues $\lambda_i$.

For example,  for Itakura-Saito distance defined as{\footnote{It is worth to note that we can generate the   large class of divergences or cost functions using Csisz\'ar  $f$-functions \cite{NMTF-09,Osterreicher02,CichZd_ICA06}.}}
\be
D_{IS}((\bp \, || \, \bq) &=& \displaystyle \sum_{i}
    \left(\frac{p_{i}}{q_{i}} +\log \frac{q_{i}}{p_{i}}  -1 \right).
    \ee
we replace ratios as follows $p_i/q_i \rightarrow \lambda_i$ and
$q_i/p_i \rightarrow \lambda^{-1}_i$, we obtain log-det divergence for SPD
\be
D_{IS}(\bP \, || \, \bQ) &=&
 \sum_{i=1}^n \left(\lambda_{i} - \log(\lambda_{i}) \right)-n,
 \ee
 which is consistent in our previous considerations (see (\ref{AlphaLD1}) and (\ref{BetaLD1})).

 As another example let consider discrete Gamma divergence defined as \cite{Cich-Amari-entropy}, \cite{Cich-Cruces-Am}
 \begin{eqnarray}
    D^{(\alpha,\beta)}_{AC}({\bp} \| {\bq})
    &=& \frac{1}{\beta(\alpha+\beta)} \log \left(\sum_{i} p_{i}^{\alpha+\beta} \right) +
    \frac{1}{\alpha(\alpha+\beta)} \log \left(\sum_{i} q_{i}^{\alpha+\beta} \right) -\frac{1}{\alpha\beta} \ln\left(\sum_{i} p_{i}^\alpha q_{i}^\beta \right) \nonumber
    \\&=&
   \frac{1}{\alpha \beta (\alpha+\beta)} \log
    \frac{\left( \displaystyle\sum_{i} p_{i}^{\alpha+\beta}\right)^{\alpha}
          \left( \displaystyle\sum_{i} q_{i}^{\alpha+\beta}\right)^{\beta}}{\left( \displaystyle\sum_{i} p_{i}^\alpha q_{i}^{\beta} \right)^{\alpha+\beta} } \\ && \mbox{for} \quad
          \alpha\neq0,\;\beta\neq0,\; \alpha+\beta \neq 0, \nonumber
          \label{divDAC11}
\end{eqnarray}
which simplifies for $\alpha=1$ and $\beta \rightarrow -1$ to the following form \cite{Cich-Amari-entropy}
\be
\lim_{\beta\rightarrow -1}D^{(1,\beta)}_{AC} (\bp \, || \, \bq) =
\displaystyle \frac{1}{n} \sum_{i=1}^n \left (\log \frac{q_i}{p_i}\right)
+ \log \left( \sum_{i=1}^n \frac{ p_i}{ q_i}\right) - \log (n) =
\log \frac{\displaystyle \frac{1}{n} \displaystyle \sum_{i=1}^n \frac{p_i}{q_i}}
{\left(\displaystyle \prod_{i=1}^n \frac{p_i}{q_i}\right)^{1/n}}.
\ee
 Hence, by substituting $p_i/q_i \rightarrow \lambda_i$, we can derive a new Gamma matrix divergence for SPD matrices:
 \be
D^{(1,0)}_{CCA} (\bP \, || \, \bQ)&=&D^{(1,-1)}_{AC} (\bP \, || \, \bQ) =
\displaystyle \frac{1}{n} \sum_{i=1}^n \left (\log \lambda^{-1}_i \right)
+ \log \left( \sum_{i=1}^n  \lambda_i\right) - \log (n) \nonumber \\
&=&
\log \frac{\displaystyle \frac{1}{n} \displaystyle \sum_{i=1}^n \lambda_i}
{\left(\displaystyle \prod_{i=1}^n \lambda_i\right)^{1/n}} =
\log \frac{M_1\{ \lambda_i\}}{M_0\{\ \lambda_i\}},
\label{LDArt-geom}
\ee
where $M_1$ denotes arithmetic means, while $M_0$ is the geometric means.

It is interesting to note that (\ref{LDArt-geom}) can be expressed equivalently as
\be
D^{(1,0)}_{CCA} (\bP \, || \, \bQ)= \log (\tr (\bP \bQ^{-1})) - \frac{1}{n} \log \det(\bP \bQ^{-1}) -\log (n).
\label{LDArt-geom2}
\ee

Similarly,  using symmetric gamma divergence defined as \cite{Cich-Amari-entropy}, \cite{Cich-Cruces-Am}:
  \begin{eqnarray}
    D^{(\alpha,\beta)}_{ACS}({\bp} \| {\bq}) &=&
   \frac{1}{ \alpha \beta } \log
    \frac{\left( \displaystyle\sum_{i} p_{i}^{\alpha+\beta}\right)
          \left( \displaystyle\sum_{i} q_{i}^{\alpha+\beta}\right)}{\left( \displaystyle\sum_{i} p_{i}^\alpha q_{i}^{\beta} \right) \left( \displaystyle\sum_{i} p_{i}^\beta q_{i}^{\alpha} \right) } \\ && \mbox{for} \quad
          \alpha\neq0,\;\beta\neq0,\; \alpha+\beta \neq 0, \nonumber
          \label{divDACS11}
\end{eqnarray}
 for  $\alpha=1$ and $\beta \rightarrow -1$, we obtain a new Gamma matrix divergence (by substituting the ratios $p_i/q_i$ by $\lambda_i$)  as follows:
\be
D^{(1,-1)}_{ACS} (\bP \, || \, \bQ) & =&
\displaystyle
\log \left( (\sum_{i=1}^n  \lambda_i)  (\sum_{i=1}^n \lambda^{-1}_i) \right) - \log (n)^2 \nonumber \\
 &= & \log \left( ( \frac{1}{n}\sum_{i=1}^n  \lambda_i) ( \frac{1}{n}\sum_{i=1}^n \lambda^{-1}_i ) \right) \nonumber \\
 &=& \log \left( M_1 \displaystyle \left\{\lambda_i \right\}  \, \displaystyle  M_1 \left\{\lambda^{-1}_i \right\} \right) \\
  &=& \log   \displaystyle \frac{ M_1\left\{\lambda_i \right\} }{ M_{-1} \left\{\lambda_i \right\}},
\label{LDArt-harm}
\ee
where $ M_{-1}\left\{\lambda_i \right\}$ denotes harmonic means.
%
Note that for $n \rightarrow \infty$  so formulated divergence can be expressed compactly as
 \be
 D^{(1,-1)}_{ACS} (\bP \, || \, \bQ) =
\log (E\{\bu \}  \; E\{\bu^{-1}\}),
\ee
where $u_i = \{\lambda_i\}$ and $u_i^{-1} = \{\lambda^{-1}_i\}$.

 The  basic  means can be defined follows:
\be
    M_\gamma({\mbi \lambda})&=&
    \left\{
    \begin{tabular}{ll}
    $M_{-\infty}=\min\{\lambda_1,\ldots,\lambda_n\}$, & $\gamma\rightarrow -\infty$,\\
     $M_{-1}= n  \left( \displaystyle\sum_{i=1}^n \displaystyle \frac{1}{\lambda_i}\right)^{-1}$, &$\gamma=-1$,\\
    $ M_{0}=\left(\displaystyle \prod_{i=1}^n \lambda_i\right)^{1/n},$& $\gamma= 0$,\\
    $  M_{1}=\displaystyle \frac{1}{n}\displaystyle \sum_{i=1}^n  \lambda_i$, &$\gamma=1$,\\
    $ M_{2}=  \left(\displaystyle \frac{1}{n} \displaystyle \sum_{i=1}^n  \lambda_i^2\right)^{1/2}$,
    & $\gamma= 2$,\\
    $M_{\infty}=\max\{\lambda_1,\ldots,\lambda_n\}$, & $\gamma\rightarrow \infty$.
    \end{tabular}
    \right.
\ee
with the following relationships between them
\be
M_{-\infty}\leq M_{-1} \leq M_{0} \leq M_{1} \leq M_{2}\leq M_{\infty},
\ee
where equalities  only holds if all $\lambda_i$ have the same values.
%
%
By increasing  the values of $\gamma$, we puts more emphasis on large relative errors that is $\lambda_i$, which are more deviated from one.
Depending on the value of $\gamma$, we obtain as particular cases: the minimum of the vector ${\mbi \lambda}$ (for $\gamma\rightarrow -\infty$), its  harmonic mean ($\gamma= -1$), the geometric mean ($\gamma= 0$), the arithmetic mean ($\gamma= 1$), the  quadratic mean ($\gamma= 2$) and the maximum of the vector ($\gamma \rightarrow -\infty$).

Exploiting the above inequalities for the means the divergence
(\ref{LDArt-geom}) and  (\ref{LDArt-harm}) can be heuristically  generalized (defined) as follows
  \be
D^{(\gamma_2,\gamma_1)}_{CCA} (\bP \, || \, \bQ)  =
\log \frac{M_{\gamma_2}\{ \lambda_i\}}{M_{\gamma_1}\{ \lambda_i\}},
\label{LDM1}
\ee
with $\gamma_2 > \gamma_1$.


The new divergence (\ref{LDM1}) is quite general and flexible and in extreme case it can take the following form:
 \be
   D^{(\infty,-\infty)}_{CCA} (\bP \, || \, \bQ) = d_{H} (\bP \, || \, \bQ) =
\log \frac{M_{\infty}\{\lambda_i\}}{M_{-\infty}\{\lambda_i\}} = \log \frac{\lambda_{max}}{\lambda_{min}},
\label{LDHilb}
\ee
which is in fact, a well-known the Hilbert projective metric \cite{Sra2014a} \cite{Reeb-Hibert-div}.


The Hilbert projective metric is extremely simple and it is suitable for big  data because it requires to compute only two (minimum and maximum) eigenvalues of  the matrix $\bP\bQ^{-1}$.

The Hilbert projective metric enjoys the following important properties \cite{Kim2014factorizations,Sra2014a}:
\begin{enumerate}

\item Nonnegativity
$d_{H} (\bP \, || \, \bQ) \geq 0$
and  Definiteness
$d_{H} (\bP \, || \, \bQ) =0$
 if and only if there is $c>0$ that $\bQ = c \bP$,

    \item Invariance to scaling
     \be
     d_{H} (c_1 \bP \, || \, c_2 \bQ) = d_{H} (\bP \, || \, \bQ)
     \ee
     for any $c_1,c_2 >0$,

\item Symmetry
\be
d_{H} (\bP \, || \, \bQ) = d_{H} (\bQ \, || \, \bP)\, .
\ee

\item Invariance under inversion
\be d_{H} (\bP \, || \,  \bQ) = d_{H} (\bP ^{-1}\, || \, \bQ^{-1})\, ,
\ee

 \item   Invariance under  congruence transformation
 \be
 d_{H} (\bA \bP \bA^{-1} \, || \, \bA \bQ \bA^{-1}) = d_{H} (\bP \, || \, \bQ)
  \ee
  for any invertible matrix $\bA$,

  \item   Invariance under  geodesic (Riemannian) transformation
 \be
 d_{H} (\bI \, || \, \bP^{-1/2} \bQ \bP^{-1/2}) = d_{H} (\bP \, || \, \bQ)\, .
 \ee

 \item Separability of divergence for the Kronecker product of SPD matrices
 \be
 d_{H} (\bP_1 \otimes \bP_2 \, || \, \bQ_1 \otimes \bQ_2) = d_{H} (\bP_1 \, || \, \bQ_1) + d_{H} (\bP_2 \, || \, \bQ_2)\, .
 \ee

 \item Scaling of power of SPD matrices
 \be
d_{H} (\bP^{\,\omega} \, || \, \bQ^{\omega})
    &=&|\omega| \; d_{H} (\bP \, || \, \bQ)
\ee
for any $\omega \neq 0 $.

Hence, for $0<|\omega_1| \leq 1\leq |\omega_2|$ we have
\be
d_{H} (\bP^{\,\omega_1} \, || \, \bQ^{\omega_1}) \leq d_{H} (\bP \, || \, \bQ)\leq d_{H} (\bP^{\,\omega_2} \, || \, \bQ^{\omega_2})\, .
\ee
 \item Scaling under weighted geometric mean
\be
d_{H} (\bP \#_s \bQ \, || \, \bP \#_u \bQ) = |s-u| \;  d_{H} (\bP \, || \, \bQ)
\ee
 for any  $ u,s \neq 0 $, where
  \be
 \bP \#_u \bQ = \bP^{1/2} (\bP^{-1/2} \bQ \bP^{-1/2})^{\;u} \; \bP^{1/2}\, .
 \ee

 \item Triangular inequality
$d_{H} (\bP \, || \, \bQ) \leq  d_{H} (\bP \, || \, \bZ) + d_{H} (\bZ \, || \, \bQ)\, .$

 \end{enumerate}

These properties can be easily derived or checked. For example, the Property
(9) can be easily derived as follows \cite{Kim2014factorizations,Sra2014a}:
\be
d_{H} (\bP \#_s \bQ \, || \, \bP \#_u \bQ) &=& d_{H} (\bP^{1/2} (\bP^{-1/2} \bQ \bP^{-1/2})^{\;s} \; \bP^{1/2} \, ||  \,(\bP^{1/2} (\bP^{-1/2} \bQ \bP^{-1/2})^{\;u} \; \bP^{1/2}) \nonumber \\
& =& d_{H} ((\bP^{-1/2} \bQ \bP^{-1/2})^{\;s} \, ||  \, (\bP^{-1/2} \bQ \bP^{-1/2})^{\;u}) \; \nonumber \\
& =& d_{H} ((\bP^{-1/2} \bQ \bP^{-1/2})^{\;(s-u)} \, ||  \, \bI) \; \nonumber \\
&=& |s-u| \;  d_{H} (\bP \, || \, \bQ)\, .
\ee

\renewcommand{\arraystretch}{2.0}
\begin{table} [p] 
\caption{Comparison of fundamental properties of 3 basic metric distances: The Riemannian (geodesic) metric (\ref{AIRM1}), Logdet Zero (Bhattacharryya) divergence (\ref{LDZ}) and the Hilbert projective metric (\ref{LDHilb}).
Matrices $\bP,\bQ,  \bP_1, \bP_2, \bQ_1, \bQ_2, \bZ \in \Real^{n \times n}$ are SPD matrices, $\bA, \bB \in \Real^{n \times n}$ are nonsingular matrices and a matrix $\bX \in \Real^{n \times r}$ with $ r <n$ is a full (column) rank matrix. The scalars satisfy the following conditions:  $c>0$, $c_1,c_2 >0$; $ 0 < \omega \leq 1$, $ s, u \neq 0$, $\psi
=|s-u|$. Geometric mean are defined $\bP \#_u \bQ = \bP^{1/2} (\bP^{-1/2} \bQ \bP^{-1/2})^{\;u} \; \bP^{1/2} $ and $\bP \#\bQ = \bP \#_{1/2}\bQ =\bP^{1/2} (\bP^{-1/2} \bQ \bP^{-1/2})^{\;1/2} \; \bP^{1/2}$.  The Hadamard product of $\bP$ and $\bQ$ is denoted by $\bP\circ\bQ$ (cf. with \cite{Sra2014a}).
\vspace{0.4cm}
}

\centering
\resizebox{1\linewidth}{!}{
\begin{tabular}{@{\extracolsep{\fill}}c||c||c}
\hline
Riemannian (geodesic) metric & LogDet Zero (Bhattacharryya) div.  & Hilbert projective metric
\\
	$d_R({\bP}\|{\bQ})= \|\log(\bQ^{-1/2} \bP \bQ^{-1/2})\|_F$
	&
 $d_{Bh}({ \bP }\|{\bQ })
  = 2 \sqrt{\log \displaystyle \frac{\det \frac{1}{2} (\bP +\bQ)}{\sqrt{\det(\bP) \det(\bQ)}}}$
	&
	$d_{H} (\bP \, \| \, \bQ) = \log
	\displaystyle \frac{\lambda_{max}\{\bP\bQ^{-1}\}}{\lambda_{min}\{\bP\bQ^{-1}\}}$
\\
\hline
$d_R(\bP\parallel\bQ)= d_R(\bQ\parallel\bP)$ &
$d_{Bh}( \bP  \parallel \bQ)= d_{Bh}(\bQ\parallel\bP)$ & $d_H(\bP \parallel\bQ)= d_H(\bQ\parallel\bP)$   \\
\hline
$d_R(c \bP \parallel c\bQ)= d_R(\bP\parallel\bQ)$ &
$d_{Bh}(c \bP\parallel c\bQ)= d_{Bh}(\bP\parallel\bQ)$ &$d_H(c_1 \bP\parallel c_2 \bQ)=d_H(\bP\parallel\bQ)$   \\
$d_R(\bA \bP\bB \parallel \bA \bQ\bB)= d_R(\bP\parallel\bQ)$ &
$d_{Bh}(\bA \bP \bB \parallel \bA \bQ\bB)= d_{Bh}(\bP\parallel\bQ)$ & $d_H(\bA\bP\bB\parallel\bA\bQ\bB)= d_H(\bP\parallel\bQ)$   \\
$d_R(\bP^{-1}\parallel\bQ^{-1})= d_R(\bP\parallel\bQ)$ &
$d_{Bh}(\bP^{-1}\parallel\bQ^{-1})= d_{Bh}(\bP\parallel\bQ)$ &$d_H(\bP^{-1}\parallel\bQ^{-1})=d_H(\bP\parallel\bQ)$   \\
\hline
$ d_R(\bP^\omega\parallel \bQ^\omega) \leq \omega \; d_R(\bP\parallel\bQ)$ &
$d_{Bh}(\bP^\omega \parallel\bQ^\omega)\leq \sqrt{\omega} \; d_{Bh}(\bP\parallel\bQ)$  & $d_{H}(\bP^\omega\parallel\bQ^\omega) \leq \omega \; d_{H}(\bP\parallel\bQ)$  \\
\hline
$d_R(\bP\parallel\bP\#_{\omega}\bQ)=\omega \; d_R(\bP\parallel\bQ)$ & $ d_{Bh}(\bP\parallel\bP\#_{\omega}\bQ)\leq \sqrt{\omega} \; d_{Bh}(\bP\parallel\bQ)$  & $d_{H}(\bP\parallel\bP\#_{\omega}\bQ)= {\omega} \; d_{H}(\bP\parallel\bQ)$  \\
$ d_R(\bZ\#_{\omega}\bP\parallel\bZ\#_{\omega}\bQ)\leq {\omega} \; d_R(\bP \parallel\bQ)$ & $d_{Bh}(\bZ\#_{\omega}\bP\parallel\bZ\#_{\omega}\bQ)\leq \sqrt{\omega} \; d_{Bh}(\bP\parallel\bQ)$  & $d_H(\bZ\#_{\omega}\bP\parallel\bZ\#_{\omega}\bQ) \leq {\omega} \; d_H(\bP \parallel \bQ)$ \\
$ d_{R} (\bP \#_s \bQ \, || \, \bP \#_u \bQ) = \psi \;  d_{R} (\bP \, || \, \bQ))$ & $d_{Bh} (\bP \#_s \bQ \, || \, \bP \#_u \bQ) \leq \sqrt{\psi} \;  d_{Bh} (\bP \, || \, \bQ)$  & $d_{H} (\bP \#_s \bQ \, || \, \bP \#_u \bQ) = \psi \; d_{H} (\bP \, || \, \bQ)$  \\
$d_R(\bP\parallel\bP\#\bQ)=d_R(\bQ\parallel\bP\#\bQ)$ & $d_{Bh}(\bP\parallel\bP\#\bQ)=d_{Bh}(\bQ \parallel \bP\#\bQ)$ & $d_H(\bP\parallel\bP\#\bQ)=d_H(\bQ\parallel\bP\#\bQ)$  \\
%
\hline
$d_R(\bZ+\bP\parallel\bZ+\bQ)\leq d_R(\bP\parallel\bQ)$ &
$d_{Bh}(\bZ+\bP\parallel\bZ+\bQ)\leq d_{Bh}(\bP,\bQ)$ &
$d_H(\bZ+\bP\parallel\bZ+\bQ) \leq d_H(\bP\parallel\bQ)$   \\
\hline
$d_R(\bX^T \bP\bX \parallel \bX^T \bQ\bX)\leq d_R(\bP\parallel\bQ)$ &
$d_{Bh}(\bX^T \bP\bX \parallel \bX^T \bQ\bX)\leq d_{Bh}(\bP\parallel\bQ)$  &
$d_H(\bX^T \bP\bX \parallel \bX^T \bQ\bX)\leq d_H(\bP\parallel\bQ)$  \\
\hline
$d_R(\bZ\otimes\bP \parallel \bZ\otimes\bQ)= \sqrt{n} \; d_R(\bP\parallel\bQ)$ &
$d_{Bh}(\bZ\otimes\bP \parallel \bZ\otimes\bQ)= \sqrt{n} \; d_{BH}(\bP \parallel \bQ)$  & $ d_H(\bZ\otimes\bP\parallel\bZ\otimes\bQ)= d_H(\bP\parallel\bQ)$  \\
\hline
\hspace{-3cm} $ d^2_R(\bP_1\otimes\bP_2 \parallel \bQ_1\otimes\bQ_2) =  $ &
\hspace{-3cm}  $d_{Bh}(\bP_1\otimes\bP_2 \parallel \bQ_1\otimes\bQ_2)  $  &
\hspace{-2.5cm}  $d_H(\bP_1\otimes\bP_2 \parallel \bQ_1\otimes\bQ_2)
 $  \\
 $=
 n\, d^2_R(\bP_1 \parallel \bQ_1)+n\, d^2_R(\bP_2 \parallel \bQ_2) + $ &
$ \geq d_{Bh}(\bP_1\circ \bP_2 \parallel \bQ_1\circ \bQ_2)$
  &
 $ = d_H(\bP_1 \parallel \bQ_1)+ d_H(\bP_2 \parallel \bQ_2)$
 \\
  $
2 \; \log \det(\bP_1 \bQ_1^{-1}) \; \log \det(\bP_2 \bQ_2^{-1})  $ &
 & $ $  \\
\hline
\end{tabular}}
\label{tableRSH}
\end{table}

In Table (\ref{tableRSH}) we summarized and compared some fundamental properties of three important metric distances: the Hilbert projective metric, the Riemannian metric and LogDet Zero (Bhattacharyya) distance (which is squared root of the S-divergence) (some of these properties are new, please compare with the results presented in \cite{Sra2014a,Bhatia-book09,Kim2014factorizations}).


\subsection{The AB Log-Det Divergence for Noisy and Ill-Conditioned Covariance Matrices}

In real-world signal processing and machine learning applications the SPD sampled  matrices can be strongly corrupted by noise and extremely ill conditioned.  In such cases  eigenvalues of generalized eigenvalue  (GEVD) problem  $\bP \bv_i=\lambda_i \bQ \bv_i$ can be divided into signal subspace and noise subspace. Signal subspace is usually represented by largest eigenvalues (and corresponding eigenvectors)  and noise subspace by smallest eigenvalues (and corresponding eigenvectors), which should be rejected. In other words, in evaluation of log-det divergences, we should take into account only these eigenvalues which represent signal subspace. The simplest approach is to find truncated dominant eigenvalues, by applying a suitable threshold $\tau>0$, that is a index $r \leq n$  for which $\lambda_{r+1} \leq \tau$ and perform summation, e.g. in Eq (\ref{DABlambda2}) form 1 to $r$ (instead form 1 to $n$) \cite{Kulis}. The threshold parameter $\tau$ can be selected via cross-validation.

Recent studies suggested that the real signal subspace  covariance matrices can be better represented by shrinking the eigenvalues.
For example, a popular and relatively simple method is to apply a thresholding and shrinkage rule to the all eigenvalues \cite{Josse-Sardy}:
\be
\widetilde \lambda_i = \lambda_i \max\{(1- \frac{\tau^{\gamma}}{\lambda^{\gamma}}),0 \},
\ee
where any eigenvalue smaller than the specific threshold  is set to zero and the rest eigenvalues are shrunk. Note that the smallest eigenvalues are more shrunk the largest one. For $\gamma=1$, we obtain a standard soft thresholding and   for $\gamma \rightarrow \infty$ a standard hard thresholding \cite{Donohooptimal13}.
We can estimate the optimal threshold $\tau >0$ and the parameter $\gamma >0$  using cross validation. However, a more practical and efficient method is to apply the Generalized Stein Unbiased Risk Estimate (GSURE) method even if the variance of noise is unknown (for detail please see \cite{Josse-Sardy} and references therein).

In this paper we have proposed alternative approach in which bias generated by noise is reduced by a suitable choice of parameters $\alpha$ and $\beta$ \cite{Cich-Cruces-Am}.
In other words, instead of eigenvalues $\lambda_i$ of the matrix $\bP\bQ^{-1}$ or its inverses, we can used regularized  or shrinked  eigenvalues \cite{Josse-Sardy}, \cite{Donohooptimal13},\cite{Donohooptimal14}. For example, on basis of formula
 (\ref{DABlambda2}) we can use the following shrinked eigenvalues{\footnote{It should be noted that equalities $\widetilde \lambda_i=1, \;\;\forall i$ hold only if all $\lambda_i$ of the matrix $\bP \bQ^{-1}$ are equal to one, which occurs only if $\bP=\bQ$.}}
\be \widetilde \lambda_i = \left(\frac{\alpha  \lambda_i^{\beta}  + \beta  \lambda_i^{-\alpha}} {\alpha+\beta}\right)^{\frac{1}{\alpha\beta}} \geq 1, \;\; \mbox{for} \;\; \alpha,\beta \neq0, \;\; \alpha,\beta >0 \;\; \mbox{or} \;\; \alpha,\beta <0,
\ee
 which play similar role to ratios $(p_i/q_i)$ (with $p_i \geq q_i$) used in the standard discrete  divergences
 \cite{Cich-Cruces-Am}, \cite{Cich-Amari-entropy}.
 So, for example, the new  gamma divergence (\ref{LDM1}) can be formulated in even more general form as
  \be
D^{(\gamma_2,\gamma_1)}_{CCA} (\bP \, || \, \bQ)  =
\log \frac{M_{\gamma_2}\{ \widetilde \lambda_i\}}{M_{\gamma_1}\{\widetilde \lambda_i\}},
\label{LDM2}
\ee
with $\gamma_2 > \gamma_1$, where $\widetilde \lambda_i$ means regularized or optimally shrinked eigenvalues.


\section{ Divergences for Multivariate Gaussian Densities  -- Differential Relative Entropies for  Multivariate Normal Distributions}

The objective of this section is to show links or relationships between family of continuous gamma divergences and AB  log-det divergences  for
multivariate Gaussian densities

Consider two multivariate Gaussian (normal) distributions:
\be
\label{MVG1}
p(\bx)&=& \displaystyle \frac{1}{\sqrt{(2 \pi)^n \det \bP}}
\exp\left(-\frac{1}{2} (\bx- \mbi \mu_1)^T \bP^{-1} (\bx- \mbi \mu_1)\right),  \\
q(\bx)&=& \displaystyle \frac{1}{\sqrt{(2 \pi)^n \det \bQ}}
\exp\left(-\frac{1}{2} (\bx- \mbi \mu_2)^T \bQ^{-1} (\bx- \mbi  \mu_2)\right), \; \bx \in \Real^n,\;\;
\label{MVG2}
\ee
where $\mbi \mu_1 \in \Real^n $ and  $\mbi \mu_2 \in \Real^n$ are means vectors and $\bP=\mbi \Sigma_1 \in \Real^{n \times n}$ and $\bQ =\mbi \Sigma_2 \in \Real^{n \times n}$ are covariance matrices of $p(\bx)$ and $q(\bx)$, respectively.

Let consider the gamma divergence for these distributions:
\begin{eqnarray}
\label{divDAC}
    D^{(\alpha,\beta)}_{AC}
    \left(p(\bx)\|q(\bx)\right)
    &=& \frac{1}{\beta(\alpha+\beta)} \log \left(\int_{\Omega} p^{\alpha+\beta}  d\bx \right) +
    \frac{1}{\alpha(\alpha+\beta)} \log \left(\int_{\Omega} q^{\alpha+\beta}  d\bx \right) -\frac{1}{\alpha\beta} \log \left(\int_{\Omega} p^\alpha q^\beta  d\bx \right) \nonumber
    \\&=&
   \frac{1}{\alpha \beta (\alpha+\beta)} \log
    \frac{\left( \displaystyle \int_{\Omega} p^{\alpha+\beta}(\bx)\; d\bx \right)^{\alpha}
          \left( \displaystyle \int_{\Omega} q^{\alpha+\beta}(\bx) \;d\bx \right)^{\beta}
          }{ \left(\displaystyle\int_{\Omega} p^\alpha(\bx) \; q^\beta(\bx) \;d\bx \right)^{\alpha+\beta}}  \\ && \mbox{for} \quad \alpha\neq0,\;\beta\neq0,\; \alpha+\beta \neq 0, \nonumber
\end{eqnarray}
 which  generalizes  a family of Gamma-divergences \cite{Cich-Cruces-Am}, \cite{Cich-Amari-entropy}.

 {\bf Theorem 3} The gamma divergence (\ref{divDAC}) for multivariate Gaussian densities (\ref{MVG1}) and (\ref{MVG2}) can be expressed in  closed form formulas as follows:
\be
\label{divDAC2}
 D^{(\alpha,\beta)}_{AC}\left(p(\bx)\|q(\bx)\right)
   &=&   \frac{1}{2}
   D_{AB}^{(\alpha,\beta)}(\bQ\|\bP)
   + \frac{1}{2(\alpha+\beta)} (\mbi \mu_1 -\mbi \mu_2)^T \left(\frac{\alpha}{\alpha+\beta} \bQ+  \frac{\beta}{\alpha+\beta}\bP \right)^{-1} (\mbi \mu_1 -\mbi \mu_2), \nonumber\\
 &=&
  \frac{1}{2\alpha\beta}
  \log \displaystyle \frac{\det\left(\displaystyle\frac{\alpha}{\alpha+\beta}\bQ+
  \frac{\beta}{\alpha+\beta}\bP\right)}
  {\det(\bQ)^\frac{\alpha}{\alpha+\beta}  \det(\bP)^\frac{\beta}{\alpha+\beta} } \\
   &&+ \frac{1}{2(\alpha+\beta)} (\mbi \mu_1 -\mbi \mu_2)^T \left(\frac{\alpha}{\alpha+\beta} \bQ+  \frac{\beta}{\alpha+\beta}\bP \right)^{-1} (\mbi \mu_1 -\mbi \mu_2), \nonumber
 \ee
 for $\alpha>0$ and $\beta>0$.

The proof of theorem is provided in the Appendix \ref{Ap-4}.

The formula (\ref{divDAC2}) consists two terms: The first term is expressed via the AB log-det divergence of the form given by (\ref{defAB5}), which is similarity between two covariance or precision matrices and is independent form the mean vectors, while the second term is a quadratic form expressed via the  Mahalanobis distance, which represents distance between means  (weighted by the covariance matrices) of the multivariate Gaussian distributions which is zero if mean values are the same.

As special important cases we obtain the following results
(some of them well-known):
\begin{enumerate}

\item For $\alpha=1$ and $\beta=0$, we obtain as the limit ($\beta \rightarrow 0$) the  Kullback-Leibler divergence can be expressed as
\cite{Davis-Dhillon2006}
\be
\label{divDAC222}
\lim_{\beta \rightarrow 0}D^{(1, \beta)}_{AC} (p(\bx) \, || \, q(\bx)) &=& D_{KL}(p (\bx) \| q (\bx)) = \int_{\Omega} p(\bx) \log \frac{p(\bx)}{q(\bx)} d\bx  \\
&=& \frac{1}{2} \left( \left(\tr(\bQ \bP^{-1}) -\log \det (\bQ \bP^{-1})-n \right) +  (\mbi \mu_1 -\mbi \mu_2)^T \bQ^{-1} (\mbi \mu_1 -\mbi \mu_2)\right), \nonumber
\ee
where the last term represents the  Mahalanobis distance, which becomes
 zero for zero-mean distributions $\mbi \mu_1 =\mbi \mu_2= 0$.

%

\item For  $\alpha=\beta=0.5$ we have the  Bhattacharyya distance \cite{Moustafa12}
\be
d_{Bh}(p \| q) &=&  -4 \log \int_{\Omega} \sqrt {p(\bx) q(\bx)} d\bx \\
&=&
2
\log \displaystyle \frac{\det \displaystyle\frac{\bP+\bQ}{2}}{\sqrt {\det \bP \det\bQ}}  + \frac{1}{2} (\mbi \mu_1 -\mbi \mu_2)^T \left[\frac{\bP+ \bQ}{2}\right]^{-1} (\mbi \mu_1 -\mbi \mu_2), \nonumber
\ee
\item For $\alpha+\beta=1$ and $0 < \alpha <1$,    we obtain the closed form expression  for the R\'enyi divergence expressed as \cite{Burbea82}
\be
D_{A}(p \| q) &=& - \frac{1}{\alpha(1-\alpha)} \log \int_{\Omega} p^{\; \alpha}(\bx) \; q^{\; 1-\alpha}(\bx) d\bx  \\
&=& \frac{1}{2
\alpha(1-\alpha)} \log \frac{\det(\alpha \bQ + (1-\alpha) \bP )}{\det (\bQ^{\alpha} \; \bP^{1-\alpha})}  + \frac{1}{2} (\mbi \mu_1 -\mbi \mu_2)^T \left[\alpha\bQ +(1-\alpha) \bP\right]^{-1} (\mbi \mu_1 -\mbi \mu_2). \nonumber
\ee

\item For $\alpha=\beta=1$, the  Gamma-divergences is reduced to the Cauchy-Schwartz divergence:
\be
D_{CS} (p(\bx) \, || \, q(\bx))
 &=&  -\log
 \displaystyle  \frac {\displaystyle  \int  p(\bx) \; q(\bx) \; d \mu(\bx) }{\left(\displaystyle  \int  p^{2}(\bx) d \mu(\bx) \right)^{1/2} \left(\displaystyle  \int  q^{2} (\bx) d \mu(\bx) \right)^{1/2}} \\
 &=&
 \frac{1}{2} \log \displaystyle \frac{\det
 \displaystyle\frac{
 (\bP^2 +\bQ^2)
 }{2}}{\det \bQ \det\bP}
   + \frac{1}{4} (\mbi \mu_1 -\mbi \mu_2)^T \left(\frac{\bP+  \bQ}{2} \right)^{-1} (\mbi \mu_1 -\mbi \mu_2)\, . \nonumber
\label{CSdiv}
\ee

\end{enumerate}

Similar formula can be derived for symmetric gamma divergence for two multivariate Gaussian. Furthermore, similar formulas can be probably derived for Elliptical Gamma distributions (EGD) \cite{Hosseini14},
which  offers more flexible modeling than the standard multivariate Gaussian distributions.

\subsection{Multiway divergences for Multivariate Normal Distributions with Separable Covariance Matrices}

Recently has been growing interest in the analysis of tensors or multiway arrays \cite{Manceur-Dutilleul,Akdemir-Gupta,Hoff-sep,Gerard-Hoff14}.
For multiway  arrays  we often use multilinear (called also array or tensor) normal distributions which correspond to the multivariate normal (Gaussian) distributions
 (\ref{MVG1})-(\ref{MVG2}), with common mean $(\mbi \mu_1 =\mbi \mu_2)$ and separable (Kronecker structured) covariance matrices expressed as\footnote{One of the most important applications of  the multilinear distributions, and hence multiway tensor analysis, is perhaps magnetic resonance imaging (MRI) (see
\cite{Ohlson13} and references therein).}}:
 \be
\bar\bP &=&  \sigma^2_P \; (\bP_1 \otimes \bP_2 \otimes \cdots \otimes \bP_K ) \in \Real^{N \times N}  \\
\bar\bQ &=&  \sigma^2_Q \; (\bQ_1 \otimes \bQ_2 \otimes \cdots \otimes \bQ_K) \in \Real^{N \times N},
\ee
where $\bP_k \in \Real^{n_k \times n_k}$ and $\bQ_k \in \Real^{n_k \times n_k}$ for $k=1,2,\ldots,K$ are SPD  matrices, usually normalized that $\det \bP_k=\det \bQ_k=1$ for each $k$ \cite{Gerard-Hoff14} and  $N = \prod_{k=1}^K n_k$.

A main advantage of the separable Kronecker model is a significant reduction in the number of variance-covariance parameters \cite{Manceur-Dutilleul}.
Usually, such separable covariance matrices are sparse and very large-scale. The challenge is to design for big data an efficient and relatively simple dissimilarity measures between two zero-mean multivariate (or multilinear) normal distributions (\ref{MVG1})-(\ref{MVG2}).
It seems that the Hilbert projective metric due to its unique properties  is a good candidate since for  the separable Kronecker structured covariances, since  it can be expressed in very simple form as:
\be
D_H(\bar\bP \parallel \bar\bQ) = \sum_{k=1}^K D_H(\bP_k \parallel \bQ_k)= \sum_{k=1}^K \log \frac{\widetilde \lambda^{(k)}_{max}}{\widetilde \lambda^{(k)}_{min}} = \log \prod_{k=1}^K \left(\frac{\widetilde \lambda^{(k)}_{max}}{\widetilde \lambda^{(k)}_{min}}\right),
\ee
where $\widetilde \lambda^{(k)}_{max}$ and $\widetilde \lambda^{(k)}_{min}$ are (shrinked) maximum and minimum eigenvalues of the (relatively small) matrices $\bP_k \bQ_k^{-1}$ for $k=1,2,\ldots,K$, respectively.  We refer to this divergence as the multiway Hilbert metric which has many attractive properties, especially invariance under multilinear transformation.

Using fundamental properties of divergence and SPD matrices we can drive other multiway log-det divergence. For example, we can obtain  the multiway Stein's loss as
\be
\hspace{-.8cm} D_{MSL}(\bar\bQ,\bar\bP)&=&
2\, D_{KL}(p(\bx) \parallel q(\bx))
=
	D^{(1,0)}_{AB}(\bar\bQ \parallel \bar\bP)\\
&=& \displaystyle 
     \tr \left({\bar \bP} {\bar \bQ}^{-1}\right) - \log \det({\bar \bP} {\bar \bQ}^{-1}) - N \nonumber \\
&=&\frac{\sigma^2_P}{\sigma^2_Q} \left(\prod_{k=1}^K  \tr(\bP_k \bQ_k^{-1}) \right)
-\sum_{k=1}^K \frac{N}{n_k} \log \det(\bP_k \bQ_k^{-1})
-N \log \left(\frac{\sigma^2_P}{\sigma^2_Q}\right)- N,
\ee
Note that under the constraints $\det \bP_k=\det \bQ_k=1$, it simplifies to
\be
D_{MSL}(\bar\bQ \parallel \bar\bP)
&=& \displaystyle 
     \tr \left({\bar \bP} {\bar \bQ}^{-1}\right) - \log \det({\bar \bP} {\bar \bQ}^{-1}) - N \\
&=&\frac{\sigma^2_P}{\sigma^2_Q} \left(\prod_{k=1}^K  \tr(\bP_k \bQ_k^{-1}) \right)
-N \log \left(\frac{\sigma^2_P}{\sigma^2_Q}\right)- N\, ,\nonumber
\ee
which is different from the multiway  Stein's loss  proposed  recently by Gerard and Hoff  \cite{Gerard-Hoff14}.

Similarly, we can derive or define multiway Riemannian metric  (under constraints that
$\det \bP_k=\det \bQ_k=1$ for each $k=1,2,\ldots,K$) as follows:
\be
d^2_R(\bar\bP \parallel \bar\bQ) = N\log^2 \frac{\sigma^2_P}{\sigma^2_Q} + \sum_{k=1}^K \frac{N}{n_k} \; d^2_R(\bP_k \parallel \bQ_k )\ .
\ee

Remark: The above multiway divergences were derived using the following properties:

If eigenvalues $\{\lambda_i\}$  and $\{\theta_j\}$ are eigenvalues with corresponding eigenvectors $\{\bv_i\}$ and $\{\bu_j\}$ for SPD matrices $\bA$ and $\bB$, respectively, then $\bA \otimes \bB$ has eigenvalues $\{\lambda_i \theta_j\}$ with corresponding eigenvectors $\{\bv_i \otimes \bu_j\}$,\\
and
\be
\bar \bP \bar \bQ^{-1} &=& (\bP_1 \otimes \bP_2 \otimes \cdots \otimes \bP_K)(\bQ^{-1}_1 \otimes \bQ^{-1}_2 \otimes \cdots \otimes \bQ^{-1}_K) \nonumber \\
&=& \bP_1\bQ^{-1}_1 \otimes \bP_2 \bQ^{-1}_2 \otimes \cdots \otimes \bP_K\bQ^{-1}_K,\\
\tr (\bar \bP \bar \bQ^{-1}) &=&\tr (\bP_1\bQ^{-1}_1 \otimes \bP_2 \bQ^{-1}_2 \otimes \cdots \otimes \bP_K\bQ^{-1}_K)= \prod_{k=1}^K \tr (\bP_k \bQ_k^{-1}),\\
\det (\bar \bP \bar \bQ^{-1}) &=& \det(\bP_1\bQ^{-1}_1 \otimes \bP_2 \bQ^{-1}_2 \otimes \cdots \otimes \bP_K\bQ^{-1}_K) = \prod_{k=1}^K (\det (\bP_k \bQ_k^{-1}))^{N/n_k}.
\ee

Other  possible extensions of  AB and Gamma matrix divergences  to separable multiway divergences for multilinear normal distributions under some normalization or constraints conditions will be discussed in our future publication.

\section{Conclusions}

In this paper, we presented  novel (dis)similarity measures: Alpha-Beta and Gamma Log-det divergences (and/or their square-roots), that smoothly connects or unifies a wide class of existing  divergences for symmetric positive definite matrices. We derived numerous results that uncovered or unified theoretic properties and qualitative  similarities between well-known divergences and also new divergences.
The scope of the results presented in this paper is vast, since the
parameterized  Alpha-Beta and Gamma log-det divergences  functions  include  several efficient and useful divergences including those based on the relative entropies, Riemannian metric (AIRM), S-divergence,  generalized Jeffreys KL or the KLDM,  Stein's loss and  Hilbert projective metric. Various links and relationships between various divergences ware also established. Furthermore, we proposed several multiway divergences for tensor (array) normal distributions.

%
%
%
%


\section{APPENDICES}

\subsection{Extension of $D^{(\alpha,\beta)}_{AB}({\bP}\|{\bQ})$ for $(\alpha,\beta)\in \mathbb{R}^2$}\label{Ap-1}
{\bf Remark}: The function (\ref{defAB1}) is only well defined in the first and third quadrant of the $(\alpha,\beta)$-plane. Outside these regions,  when parameters and $\alpha$ and $\beta$ have opposite signs (i.e. $\alpha >0$ and $\beta <0$ or vice versa $\alpha < 0$ and $\beta >0$), the divergence can be complex valued. This undesired behavior can be avoided with the help of the truncation operator
\be
[x]_+ =
    \left\{
    \begin{array}{ll}
    x & x\geq 0\\
    0, & x<0,
    \end{array}
    \right.
\ee
that will be used to prevent the arguments of the logarithms to be negative. The new definition of the AB log-det divergence
\begin{eqnarray}\label{defAB1a}
    D^{(\alpha,\beta)}_{AB}({\bP}\|{\bQ})
    &=&
    \frac{1}{\alpha \beta} \log \left[\det
     \frac{\alpha (\bP \bQ^{-1})^{\beta} + \beta (\bP \bQ^{-1})^{-\alpha}}
        {\alpha+\beta}\right]_{+} \\ \nonumber
    \\&&\ \text{for}\ \alpha \neq 0, \;\; \beta\neq 0, \;\;\; \alpha+\beta\neq 0. \nonumber
\end{eqnarray}
is compatible with the previous one on the first and third quadrant of the $(\alpha,\beta)$ plane, while it is also well defined on the second and four quadrants except for the special cases $ \alpha =0,\  \beta= 0,\  \alpha+\beta= 0$ where the formula is undetermined. Enforcing the continuity, we can define explicitly the AB-log-det divergence on the entire $(\alpha,\beta)$-plane as:
\be
    \label{ABdef-full1}
    D^{(\alpha,\beta)}_{AB}({\bP}\|{\bQ})
    \!\!\!&=&\!\!\!
    \left\{
    \begin{tabular}{ll}
    $\displaystyle  \frac{1}{\alpha \beta} \log \det
    \left[
     \frac{\alpha (\bP \bQ^{-1})^{\beta} + \beta (\bQ \bP^{-1})^{\alpha}}
        {\alpha+\beta}
        \right]_+
        $&  $\mbox{for} \;\; \alpha,\beta\neq0, \;\;  \alpha+\beta \neq 0$ \\
    \\
    $ \displaystyle \frac{1}{\alpha^2} \left[
     \tr \left((\bQ \bP^{-1})^{\alpha} - \bI\right) - \alpha \log \det (\bQ \bP^{-1}) \right]
         $   &$\ \text{for}\ \;\; \alpha \neq 0, \; \beta= 0 $ \\
    \\
    $\displaystyle  \frac{1}{\beta^2} \left[
     \tr \left((\bP \bQ^{-1})^{\beta} - \bI \right) - \beta \log \det (\bP \bQ^{-1}) \right] $   &$\
      \text{for}\ \;\; \alpha= 0, \; \beta\neq 0 $ \\
       \\
      $\displaystyle \frac{1}{\alpha^2}
      \displaystyle \log \det [(\bP \bQ^{-1})^{-\alpha}(\bI +\log(\bP\bQ^{-1})^\alpha)]_+	 ^{-1}	
      $
      &  
     $\ \text{for}\  \alpha=-\beta$ \\
      \\
    $\displaystyle \frac{1}{2} \tr \log^2 (\bP \bQ^{-1}) = \frac{1}{2} ||\log(\bQ^{-1/2} \bP \bQ^{-1/2})||^2_F$ & $\ \text{for}\ \;\; \alpha, \; \beta= 0 $.
    \end{tabular}
    \right.
\ee

\subsection{Domain of the eigenvalues for which $D^{(\alpha,\beta)}_{AB}({\bP}\|{\bQ})$ is  finite}\label{Ap-2}
 In this section, we assume that $\lambda_i$, the eigenvalues of ${\bP}{\bQ}^{-1}$, satisfy that $0\leq\lambda_i\leq \infty$ for all~$i=1,\ldots,n$. We will determine the bounds on the eigenvalues of ${\bP}{\bQ}^{-1}$ that prevent the AB log-det  divergence to be infinite. For this purpose, let us recall that
\be
 D^{(\alpha,\beta)}_{AB}({\bP}\|{\bQ})
     &=&  \frac{1}{\alpha \beta} \sum_{i=1}^n \log \left[\frac
     {\alpha \lambda_i^{\beta}  + \beta \lambda_i^{-\alpha}}
        {\alpha+\beta}\right]_+, \;\; \alpha,\;\beta,\; \alpha+\beta \neq 0.
        \label{DABlambda22}
\ee
Let us assume that $0\leq\lambda_i\leq \infty$ for all $i$. For the divergence to be finite, the arguments of the logarithms in the previous expression should be all positive. This happens for
\be
\frac{\alpha \lambda_i^{\beta}  + \beta \lambda_i^{-\alpha}}{\alpha+\beta}>0\qquad  \forall i,
\ee
condition which is always true when $\alpha,\beta>0$ or when $\alpha, \beta<0$. On the contrary, when $\text{sign}(\alpha\beta)=-1$, we have the following two cases. On the one hand, for $\alpha>0$, we can solve initially for $\lambda_i^{\alpha+\beta}$ and later for $\lambda_i$ to obtain
\be
\frac{\lambda_i^{\alpha+\beta}}{\alpha+\beta}  > \frac{-\beta}{\alpha (\alpha+\beta)}=\left|\frac{\beta}{\alpha}\right|\frac{1}{\alpha+\beta}
\quad \longrightarrow \quad
\lambda_i > \left|\frac{\beta}{\alpha}\right|^\frac{1}{\alpha+\beta} \qquad \forall i,\  \text{for}\ \alpha>0\ \text{and}\ \beta<0.
\ee
On the other hand, for $\alpha<0$, we obtain
\be
\frac{\lambda_i^{\alpha+\beta}}{\alpha+\beta}  < \frac{-\beta}{\alpha (\alpha+\beta)}=\left|\frac{\beta}{\alpha}\right|\frac{1}{\alpha+\beta}
\quad \longrightarrow \quad
\lambda_i < \left|\frac{\beta}{\alpha}\right|^\frac{1}{\alpha+\beta} \qquad \forall i,\  \text{for}\ \alpha<0\ \text{and}\ \beta>0.
\ee

$\text{sign}(\alpha\beta)=-1$, we can solve for $\lambda_i^{\alpha+\beta}$ to obtain
\be
\frac{\lambda_i^{\alpha+\beta}}{\alpha+\beta} &>& \left|\frac{\beta}{\alpha}\right|\frac{1}{\alpha+\beta}  \qquad  \forall i.
\ee
Solving again for $\lambda_i$ we see that
\be
\lambda_i &>& \left|\frac{\beta}{\alpha}\right|^\frac{1}{\alpha+\beta} \qquad \forall i,\  \text{for}\ \alpha>0\ \text{and}\ \beta<0,
\ee
and
\be
\lambda_i &<& \left|\frac{\beta}{\alpha}\right|^\frac{1}{\alpha+\beta} \qquad \forall i,\  \text{for}\ \alpha<0\ \text{and}\ \beta>0.
\ee
Moreover, in the limit, when $\alpha\rightarrow -\beta\neq 0$ these bounds simplify to
\be
\lim_{\alpha\rightarrow -\beta}\left|\frac{\beta}{\alpha}\right|^\frac{1}{\alpha+\beta}=
e^{-1/\alpha}\
 \qquad \forall i,\  \text{for}\ \beta\neq 0.
\ee
Whereas, in the limit, for $\alpha\rightarrow 0$ or for $\beta\rightarrow 0$ the bounds disappear. The lower-bounds converge to $0$, while the upper-bounds converge to $\infty$, leading to the trivial inequalities $0<\lambda_i<\infty$.

This concludes the determination of the domain of the eigenvalues for which the divergence is finite. Outside of this domain we should expect that  $D^{(\alpha,\beta)}_{AB}({\bP}\|{\bQ})=\infty$. The complete picture of bounds for different values of $\alpha$ and $\beta$ is shown in Fig. \ref{Fig:contour}.

%
%
%
\subsection{Proof of the Non-negativity of $D^{(\alpha,\beta)}_{AB}({\bP}\|{\bQ})$}\label{Ap-3}
The AB log-det divergence is a separable as a sum of the individual divergences of the eigenvalues from the unity, i.e.
\be
 D^{(\alpha,\beta)}_{AB}({\bP}\|{\bQ})
     &=&   \sum_{i=1}^n D^{(\alpha,\beta)}_{AB}(\lambda_i\|1)
\ee
where
\be
 D^{(\alpha,\beta)}_{AB}(\lambda_i\|1)
     &=&  \frac{1}{\alpha \beta} \log \left[\frac
     {\alpha \lambda_i^{\beta}  + \beta \lambda_i^{-\alpha}}
        {\alpha+\beta}\right]_+, \;\; \alpha,\;\beta,\; \alpha+\beta \neq 0.
        \label{DABlambda22z1}
\ee

Then, we can prove the non-negativity of $D^{(\alpha,\beta)}_{AB}({\bP}\|{\bQ})$ just showing that the divergence on each of the eigenvalues $D^{(\alpha,\beta)}_{AB}(\lambda_i\|1)$ is non-negative and minimum at $\lambda_i=1$.

For this purpose, we first realize that the only critical point of the criterion is obtained for $\lambda_i=1$. This can be seen equating to zero the derivative of the criterion 
\be
\frac{\partial D^{(\alpha,\beta)}_{AB}(\lambda_i\|1)}{\partial \lambda_i}   	
     &=&   \frac{ \lambda_i^{\alpha+\beta}-1}
        {\alpha\lambda_i^{\alpha+\beta+1}+\beta \lambda_i}=0
        \label{DABlambda22z2}
\ee
and solving for $\lambda_i$.

Next we will show that the sign of the derivative only changes at  the critical point $\lambda_i=1$.
If we rewrite
\be
\frac{\partial D^{(\alpha,\beta)}_{AB}(\lambda_i\|1)}{\partial \lambda_i}   	
        &=&  \left(\frac{\lambda_i^{\alpha+\beta}-1}{\alpha+\beta}\right)
        \left(\lambda_i\frac{\alpha\lambda_i^{\alpha+\beta}+\beta }{\alpha+\beta}
        \right)^{-1}
        \label{DABlambda22z3}
\ee
and observe that the condition of the divergence to be finite enforces $\frac{\alpha\lambda_i^{\alpha+\beta}+\beta}{\alpha+\beta} >0$, then it follows that
\be
\sign \left\{\frac{\partial D^{(\alpha,\beta)}_{AB}(\lambda_i\|1)}{\partial \lambda_i}\right\}
\equiv
\sign \left\{\frac{\lambda_i^{\alpha+\beta}-1}{\alpha+\beta}\right\}
=
    \left\{
    \begin{array}{ll}
    -1 & \text{ for}\ \ \lambda_i<1\\
    0, & \text{ for}\ \ \lambda_i=1\\
    +1 & \text{ for}\ \ \lambda_i>1.\\
    \end{array}
    \right.
\ee
Since  the derivative is strictly negative for $\lambda_i<1$ and strictly positive for $\lambda_i>1$,  the  critical point at $\lambda_i=1$ is the global minimum of  $D^{(\alpha,\beta)}_{AB}(\lambda_i\|1)$. From this result, the non-negativity of the divergence $D^{(\alpha,\beta)}_{AB}({\bP}\|{\bQ})\geq 0$ easily follows. Moreover, $D^{(\alpha,\beta)}_{AB}({\bP}\|{\bQ})=0$ only for $\lambda_i=1$ for $i=1,\ldots,n$, which concludes the proof of the Theorem 1 and 2.

\subsection{Derivation of the Riemannian Metric (\ref{RimA}) }\label{Ap-3A}

We calculate $D_{AB}^{(\alpha,\beta)} (\bP +d \bP \parallel \bP)$ by Taylor expansion when $d \bP$ is small. From
\be
(\bP +d \bP )\bP^{-1} = \bI + d \bZ,
\ee
where
\be
d \bZ &=& d \bP \bP^{-1}, \nonumber \\
\alpha [(\bP + d \bP)\bP^{-1}]^{\beta} &=& \alpha \bI +\alpha \beta \, d \bZ + \frac{\alpha\beta (\beta-1)}{2} \; d \bZ \, d \bZ +  O (|d \bZ|^3). \nonumber
\ee
Similar calculations hold for $\beta [(\bP + d \bP)\bP^{-1}]^{-\alpha}$, and
\be
\alpha [(\bP + d \bP)\bP^{-1}]^{\beta} +\beta (\bP + d \bP)\bP^{-1}]^{-\alpha}= (\alpha+\beta)\left(\bI +  \frac{\alpha\beta }{2} d \bZ \, d \bZ\right), \nonumber
\ee
where the first-order tern of $d \bZ$ disappears and the higher-order terms are neglected. \\
Since
\be
\det \left(\bI +  \frac{\alpha\beta }{2} d \bZ \, d \bZ \right) = 1+  \frac{\alpha\beta }{2} \tr( d \bZ \, d \bZ),
\ee
by taking its logarithm, we have
\be
D_{AB}^{(\alpha,\beta)} (\bP +d \bP \parallel \bP)=\frac{1}{2} \tr (d \bP \, \bP^{-1} \, d \bP \, \bP^{-1}),
\ee
for any $\alpha$ and $\beta$.

\subsection{Gamma divergence for multivariate Gaussian densities} \label{Ap-4}
%
We start recalling that, for a given quadratic function $f(\bx)=-c+\bb^T \bx-\frac{1}{2}\bx^T \bA\bx$ where $\bA$ is a positive definite symmetric matrix, the integral of $\exp\{f(\bx)\}$ with respect to $\bx$ is given by
\be \label{Integral}
	\int_\Omega 	e^{-\frac{1}{2}\bx^T \bA\bx+\bb^T \bx-c} d\bx
	&=&
	(2\pi)^\frac{N}{2} \det(\bA)^{-\frac{1}{2}}\ e^{\frac{1}{2}\bb^T \bA^{-1}\bb-c}.
\ee
This formula has been obtained by evaluated the integral as follows
\be \label{Integralp}
	\int_\Omega 	e^{-\frac{1}{2}\bx^T \bA\bx+\bb^T \bx-c} d\bx
	&=&
	e^{\frac{1}{2}\bb^T \bA^{-1}\bb-c}
	\int_\Omega e^{-\frac{1}{2}\bx^T \bA\bx+\bb^T \bx-\frac{1}{2}\bb^T \bA^{-1}\bb} d\bx\\
	&=&
	e^{\frac{1}{2}\bb^T \bA^{-1}\bb-c}
	\int_\Omega e^{ (\bx-\bA^{-1}\bb)^T \bA (\bx-\bA^{-1}\bb)} d\bx\\
	&=&
	e^{\frac{1}{2}\bb^T \bA^{-1}\bb-c}\ (2\pi)^\frac{N}{2} \det(\bA)^{-\frac{1}{2}},
\ee
assuming that $\bA$ is symmetric positive definite matrix (which assures the convergence of the integral and  the validity of (\ref{Integral})).

The Gamma divergence involves the a product of densities that, in the multivariate Gaussian case, we can simplify as
\be
	p^\alpha(\bx) q^\beta(\bx)
	&=&
	(2\pi)^{-\frac{N}{2}(\alpha+\beta)}\det(\bP)^{-\frac{\alpha}{2}} \det(\bQ)^{-\frac{\beta}{2}}
	\times \nonumber\\
	&&\exp
	\left\{
	-\frac{\alpha}{2}(\bx -\bmu_1)^T \bP^{-1}(\bx -\bmu_1)
	-\frac{\beta}{2}(\bx -\bmu_2)^T \bQ^{-1}(\bx -\bmu_2)
	\right\}\\
	&=&
	d\
	\exp
	\left\{
	-c+\bb^T \bx-\frac{1}{2}\bx^T \bA\bx
	\right\},
\ee
where
\be
	\bA &=&  \alpha \bP^{-1}+ \beta \bQ^{-1}	
	\\
	\bb &=& \left(\bmu_1^T \alpha \bP^{-1}+\bmu_2^T \beta \bQ^{-1}\right)^T
	\\
	c &=& \frac{1}{2}\bmu_1(\alpha \bP^{-1}) \bmu_1 + \frac{1}{2}\bmu_2(\beta \bQ^{-1}) \bmu_2
	\\
	d &=& (2\pi)^{-\frac{N}{2}(\alpha+\beta)}\det(\bP)^{-\frac{\alpha}{2}} \det(\bQ)^{-\frac{\beta}{2}}.
\ee
Integrating this product with the help of (\ref{Integral}), we obtain
\be \label{Integral1}
	\int_\Omega 	p^\alpha(\bx) q^\beta(\bx)  d\bx
	&=&
	d\ (2\pi)^\frac{N}{2} \det(\bA)^{-\frac{1}{2}}\ e^{\frac{1}{2}\bb^T \bA^{-1}\bb-c}\\	 
	&=&
	(2\pi)^{\frac{N}{2}(1-(\alpha+\beta))}\det(\bP)^{-\frac{\alpha}{2}} \det(\bQ)^{-\frac{\beta}{2}}\
	\det(\alpha \bP^{-1}+ \beta \bQ^{-1}	)^{-\frac{1}{2}}\times \nonumber\\
	&& e^{\frac{1}{2}
	\left(\bmu_1^T \alpha \bP^{-1}+\bmu_2^T \beta \bQ^{-1}\right)
	(\alpha \bP^{-1}+ \beta \bQ^{-1}	)^{-1}
	\left(\bmu_1^T \alpha \bP^{-1}+\bmu_2^T \beta \bQ^{-1}\right)^T
	}\times
	\nonumber\\
	&&e^{-\frac{1}{2}\bmu_1(\alpha \bP^{-1}) \bmu_1 - \frac{1}{2}\bmu_2(\beta \bQ^{-1}) \bmu_2 },
\ee
provided that $\alpha \bP^{-1}+ \beta \bQ^{-1}$ is positive definite.

Rearranging the expression in terms of $\bmu_1$ and $\bmu_2$ gives
\be \label{Integral2}
	\int_\Omega 	p^\alpha(\bx) q^\beta(\bx)  d\bx
	&=&
	(2\pi)^{\frac{N}{2}(1-(\alpha+\beta))}\det(\bP)^{-\frac{\alpha}{2}} \det(\bQ)^{-\frac{\beta}{2}}\
	\det(\alpha \bP^{-1}+ \beta \bQ^{-1}	)^{-\frac{1}{2}}
	\times \nonumber\\
	&& e^{\frac{1}{2}
	\bmu_1^T
	\left[\alpha \bP^{-1}(\alpha \bP^{-1}+ \beta \bQ^{-1}	)^{-1}\alpha \bP^{-1}-\alpha \bP^{-1}\right]
	\bmu_1
	}\times\nonumber\\
	&&e^{\frac{1}{2}
	\bmu_2^T
	\left[\beta \bQ^{-1}(\alpha \bP^{-1}+ \beta \bQ^{-1}	)^{-1}\beta \bQ^{-1}-\alpha \bQ^{-1}\right]
	\bmu_2
	}
	\times\nonumber\\
	&& e^{
	\bmu_1^T \alpha \bP^{-1}
	(\alpha \bP^{-1}+ \beta \bQ^{-1}	)^{-1}
	\beta \bQ^{-1}\bmu_2.
	}
\ee
With the help of the Woodbury matrix identity we can simplify
\be
 	e^{\frac{1}{2}
	\bmu_1^T
	\left[\alpha \bP^{-1}(\alpha \bP^{-1}+ \beta \bQ^{-1}	)^{-1}\alpha \bP^{-1}-\alpha \bP^{-1}\right]
	\bmu_1
	}
	=
e^{-\frac{1}{2}
	\bmu_1^T
	(\alpha^{-1} \bP+ \beta^{-1} \bQ )^{-1}
	\bmu_1
	}		
\ee
\be
	e^{\frac{1}{2}
	\bmu_2^T
	\left[\beta \bQ^{-1}(\alpha \bP^{-1}+ \beta \bQ^{-1}	)^{-1}\beta \bQ^{-1}-\beta \bQ^{-1}\right]
	\bmu_2
	}
	=
	e^{-\frac{1}{2}
	\bmu_2^T
	(\alpha^{-1} \bP+ \beta^{-1} \bQ )^{-1}
	\bmu_2
	}
\ee
\be
	e^{
	\bmu_1^T \alpha \bP^{-1}
	(\alpha \bP^{-1}+ \beta \bQ^{-1}	)^{-1}
	\beta \bQ^{-1}\bmu_2
	}
	=
	e^{
	\bmu_1^T (\alpha^{-1} \bP+ \beta^{-1} \bQ )^{-1} \bmu_2
	}
\ee
arriving to the desired result:

\be \label{Integral-pq}
	\int_\Omega 	p^\alpha(\bx) q^\beta(\bx)  d\bx
	&=&
	(2\pi)^{\frac{N}{2}(1-(\alpha+\beta))}\det(\bP)^{-\frac{\alpha}{2}} \det(\bQ)^{-\frac{\beta}{2}}\
	(\alpha+\beta)^{-\frac{N}{2}}\times \nonumber\\
	&& \det\left(\frac{\alpha}{\alpha+\beta} \bP^{-1}+ \frac{\beta}{\alpha+\beta} \bQ^{-1}	 \right)^{-\frac{1}{2}}	\times \nonumber\\
	&& e^{-\frac{\alpha\beta}{2(\alpha+\beta)}(\bmu_1-\bmu_2)^T\left(\frac{\beta}{\alpha+\beta} \bP+ \frac{\alpha}{\alpha+\beta} \bQ \right)^{-1}(\bmu_1-\bmu_2).
	}
\ee
This formula can be can easily particularized to evaluate the integrals
\be \label{Integral-p}
	\int_\Omega 	p^{\alpha+\beta}(\bx)  d\bx
	&=&
	\int_\Omega 	p^\alpha(\bx) p^\beta(\bx)  d\bx \nonumber \\
	&=&
	(2\pi)^{\frac{N}{2}(1-(\alpha+\beta))}\det(\bP)^{-\frac{\alpha}{2}} \det(\bP)^{-\frac{\beta}{2}}\
	\det(\alpha \bP^{-1}+ \beta \bP^{-1}	)^{-\frac{1}{2}}
	\times \nonumber\\
	&& e^{-\frac{\alpha\beta}{2(\alpha+\beta)}(\bmu_1-\bmu_1)^T\left(\frac{\beta}{\alpha+\beta} \bP+ \frac{\alpha}{\alpha+\beta} \bP \right)^{-1}(\bmu_1-\bmu_1)
	}
	\nonumber \\
	&=&
	 (2\pi)^{\frac{N}{2}(1-(\alpha+\beta))}(\alpha+\beta)^{-\frac{N}{2}}\det(\bP)^{\frac{1-(\alpha+\beta)}{2}}
\ee
and
\be \label{Integral-p1}
	\int_\Omega 	q^{\alpha+\beta}(\bx)  d\bx
	&=& (2\pi)^{\frac{N}{2}(1-(\alpha+\beta))}
(\alpha+\beta)^{-\frac{N}{2}}\det(\bQ)^{\frac{1-(\alpha+\beta)}{2}}.
\ee

By substituting these integrals into the definition of the gamma divergence and simplifying, we obtain generalized closed form formula:
\be
 D^{(\alpha,\beta)}_{AC}\left(p(\bx)\|q(\bx)\right)
 &=&
   \frac{1}{\alpha \beta } \log
    \frac{\left( \displaystyle \int_{\Omega} p^{\alpha+\beta}(\bx)\; d\bx \right)^\frac{\alpha}{\alpha+\beta}
          \left( \displaystyle \int_{\Omega} q^{\alpha+\beta}(\bx) \;d\bx \right)^\frac{\beta}{\alpha+\beta}
          }{ \displaystyle\int_{\Omega} p^\alpha(\bx) \; q^\beta(\bx) \;d\bx }
          \nonumber	\\
  &=&
  \frac{1}{2\alpha\beta}
  \log \displaystyle \frac{\det\left(\displaystyle\frac{\alpha}{\alpha+\beta}\bQ+
  \frac{\beta}{\alpha+\beta}\bP\right)}
  {\det(\bQ)^\frac{\alpha}{\alpha+\beta}  \det(\bP)^\frac{\beta}{\alpha+\beta} } \\
   &&+ \frac{1}{2(\alpha+\beta)} (\mbi \mu_1 -\mbi \mu_2)^T \left(\frac{\alpha}{\alpha+\beta} \bQ+  \frac{\beta}{\alpha+\beta}\bP \right)^{-1} (\mbi \mu_1 -\mbi \mu_2), \nonumber
 \ee
which concludes the proof the Theorem 3.

\end{document}